\documentclass[12pt]{article}

\AtBeginDocument{
  \addtocontents{toc}{\normalsize}
  \addtocontents{lof}{\tiny}
}
\pdfoutput=1

\usepackage{amsmath,amssymb,amsfonts,amsthm,amscd,mathrsfs}
\usepackage{xcolor}
\definecolor{darkblue}{rgb}{0.1,0.1,.7}
\usepackage[]{version}
\usepackage{float}
\usepackage[]{graphicx}
\usepackage[]{latexsym}
\usepackage{geometry}
\usepackage{subcaption}
\usepackage[utf8]{inputenc}
\usepackage[all,cmtip]{xy}
\usepackage{empheq}

\usepackage{ifthen}
\usepackage{tikz}
\usepackage{array,setspace,yfonts,dsfont,bbm,colonequals}
\usepackage[colorlinks, linkcolor=darkblue, citecolor=darkblue, urlcolor=darkblue, linktocpage]{hyperref} 

\usepackage{cite}
\usepackage{xspace}
\usepackage{empheq}
\usepackage{xcolor}

\numberwithin{equation}{section}
\newcommand{\ba}{\begin{equation}\begin{aligned}}
\newcommand{\ea}{\end{aligned}\end{equation}}

\newcommand{\tj}{\bar \jmath_0}

\newcommand{\OU}{\Omega^B}
\newcommand{\OL}{\Omega^F}
\newcommand{\OLU}{\Omega^{B,F}}
\newcommand{\fl}{f^F}
\newcommand{\fu}{f^B}
\newcommand{\gl}{g^F}
\newcommand{\gu}{g^B}
\newcommand{\ghu}{\hat g^B}
\newcommand{\ghl}{\hat g^F}
\newcommand{\ful}{f^{B,F}}
\newcommand{\gul}{g^{B,F}}
\newcommand{\ghul}{\hat g^{B,F}}

\newcommand{\reef}[1]{(\ref{#1})}
\newcommand{\be}{\begin{equation}}
\newcommand{\ee}{\end{equation}}

\newcommand{\bea}{\begin{equation}\begin{aligned}}
\newcommand{\eea}{\end{aligned}\end{equation}}

\newcommand{\ud}{\mathrm d}

\newcommand{\Df}{{\Delta_\phi}}

\begin{document}

\vspace*{-.6in} \thispagestyle{empty}
\begin{flushright}
\end{flushright}
\vspace{1cm} {\Large
\begin{center}
{\bf Dispersion relations and exact bounds on CFT correlators}\\
\end{center}}
\vspace{1cm}
\begin{center}
{Miguel F.~Paulos}\\[1cm] 
{
\small
{\em Laboratoire de Physique de l'\'Ecole Normale Sup\'erieure\\ PSL University, CNRS, Sorbonne Universit\'es, UPMC Univ. Paris 06\\ 24 rue Lhomond, 75231 Paris Cedex 05, France}
}\normalsize
\\
\end{center}

\begin{center}
	{\texttt{miguel.paulos@ens.fr} 
	}
	\\
\end{center}

\vspace{4mm}

\begin{abstract}
We derive new crossing-symmetric dispersion formulae for CFT correlators restricted to the line. The formulae are equivalent to the sum rules implied by what we call master functionals, which are analytic extremal functionals which act on the crossing equation. The dispersion relations provide an equivalent formulation of the constraints of the Polyakov bootstrap and hence of crossing symmetry on the line. The built in positivity properties imply simple and exact lower and upper bounds on the values of general CFT correlators on the Euclidean section, which are saturated by generalized free fields. Besides bounds on correlators, we apply this technology to determine new universal constraints on the Regge limit of arbitrary CFTs and obtain very simple and accurate representations of the 3d Ising spin correlator.

\end{abstract}
\vspace{2in}


\newpage

{
\setlength{\parskip}{0.05in}
\tableofcontents
\renewcommand{\baselinestretch}{1.0}\normalsize
}


\setlength{\parskip}{0.1in}
\newpage

\section{Introduction}\label{sec:introduction}
\begin{figure}%
\begin{center}
\hspace{-2cm}
\includegraphics[width=12cm]{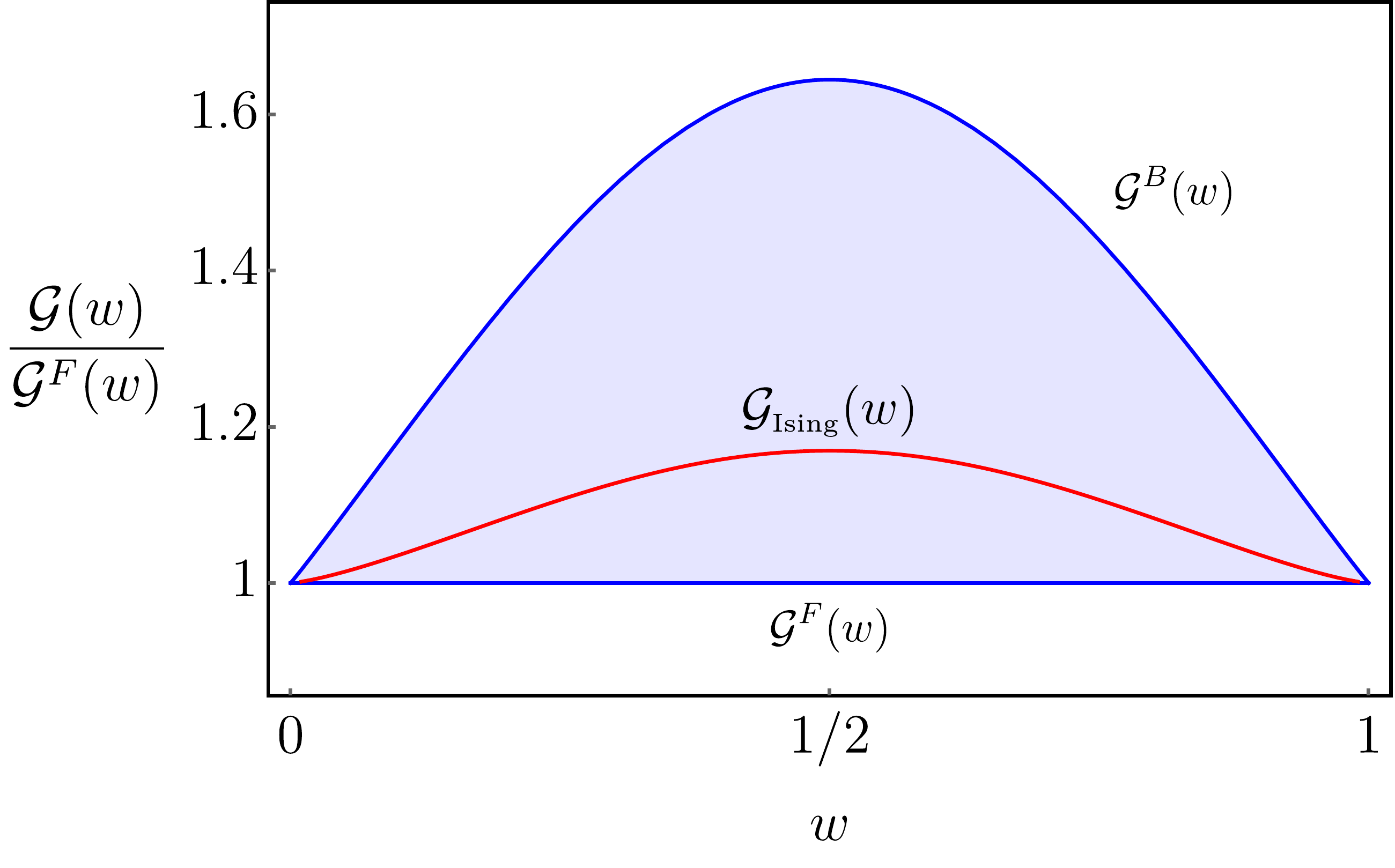}%
\caption{Upper and lower bounds on CFT correlators $\mathcal G(w):=\mathcal G(w,w)$ of a field of dimension $\Df$, shown for the case $\Df=\Delta_\sigma^{\mbox{\tiny Ising}}\sim 0.518$. Here $\mathcal G^{B,F}$ stand for the Bosonic/Fermionic generalized free field correlators, $\mathcal G^{B,F}(w)=\pm 1+w^{-2\Df}+(1-w)^{-2\Df}$. Any unitary CFT correlator must live inside the shaded region, with the caveat that the upper bound is only valid for correlators where the first operator has scaling dimension $\Delta_g\geq 2\Df$. In red the 3d Ising spin field correlator computed from the CFT data provided in \cite{Simmons-Duffin:2016wlq}.}%
\label{isingbound}%
\end{center}
\end{figure}

Are CFT correlators free to take values as they please?  Alas, the present work says no: even they are confined and allowed to wander only within a limited range, see figure \ref{isingbound}. This is yet another restriction on an ever-growing list of indignities \cite{Rattazzi:2008pe,Poland:2018epd}.

We will find an answer to this question from a study of what we call {\em master functionals}: these are special analytic functionals whose action on the crossing equation repackages its contents and provides a new window into its implications. In principle we expect there to be one such  functional for every extremal solution to crossing \cite{ElShowk:2012hu,El-Showk:2016mxr,Gliozzi2013} (i.e. solutions which saturate a bound on the CFT data). In practice, we can only construct them exactly for two very special cases, where these solutions correspond to the generalized free fermion and boson correlators in $d=1$. Each of the corresponding master functionals acts as a generating function for certain analytic functional bases acting on the crossing equation \cite{Mazac:2016qev,Mazac:2018,Mazac2019a}. The action of a given master functional on the crossing equation leads to a one-parameter family of sum rules which is equivalent to a dispersion relation for a CFT correlator.
Each such dispersion relation is crossing symmetric, which means that its action on an $s$-channel conformal block (say) will yield a fully crossing symmetric Polyakov block. As such, the statement that a correlator satisfies one of these dispersion relations is automatically equivalent to the Polyakov bootstrap\footnote{The Polyakov(-Mellin) bootstrap started in the visionary work \cite{Polyakov:1974gs}. It was modernized and made fit for consumption through \cite{Sen:2015doa,Gopakumar:2016wkt, Gopakumar:2018xqi}, and finally put on a solid footing in 1d  in \cite{Mazac2019a}. A proposal for a higher dimensional version has appeared in \cite{Caron-Huot:2020adz}. See also \cite{Penedones:2019tng,Carmi:2020ekr} as well as \cite{Kaviraj2018,Mazac:2018biw} for related works.}. The dispersion relation provides us with an efficient way to compute Polyakov blocks which makes their positivity properties manifest. These positivity properties in turn imply bounds on the allowed values of CFT correlators.

The dispersion relations express the behaviour of a CFT correlator on the complexified line $z=\bar z$ in terms of its double discontinuity, up to a finite set of low energy data. This allows us to take a plunge deep into the complex plane to explore conformal Regge kinematics \cite{Costa:2012cb}, tethered to the relative safety of the Euclidean line. We find that the Regge limit is either dominated by low energy ``subtractions'' to the correlator, or fixed via the double discontinuity by the high energy spectrum. In particular, we show that decaying interactions at high energies can be related to a sufficiently ``free'' spectrum at high energies.\footnote{A similar statement was made in \cite{Caron-Huot:2020ouj} where a beautiful (conjectural) connection was also made to the behaviour of leading Regge trajectory in the spectrum.}

This work focuses on a subset of the full constraints of crossing symmetry, where we study CFT correlators where all operators lie on a (complexified) line. We will also restrict mostly our analysis to take into account only the existence of an $SL(2,\mathbb R)$ subgroup of the full conformal group. This restriction erases the information about spin in higher dimensions. Nevertheless the resulting constraints are still incredibly strong, leading to non-trivial consequences even for generic CFTs. We emphasize that: 

\begin{center}
{\em All results presented in this work must hold for generic CFTs in any spacetime dimension  i.e. they are {\em not} restricted to 1d CFTs.}
\end{center}
We should mention that our results bear a close relationship to recent work considering dispersion relations and associated sum rules for CFT correlators, taking into acccount their full cross-ratio dependence and in particular the dynamics of spin \cite{Carmi:2019cub,Mazac:2019shk,Caron-Huot:2020adz}. We will comment on the similarities and differences between this work and those in the last section of this paper.

We now turn to the outline of this work and some of the main results.

\subsection{Outline and summary of results}
After a brief review of 1d CFT kinematics and extremal functional bases in section \ref{sec:review}, we begin in earnest by introducing in section \ref{sec:master} the general problem of constraining the allowed values of CFT correlators on the line $z=\bar z$. A solution to this problem motivates the introduction of master functionals, which act as generating functions for 1d functional bases:
\bea
\OL_w&=-\sum_{n=0}^{+\infty} \left[G_{\Delta_n^F}(w|\Df) \alpha_n^F+\partial_\Delta G_{\Delta_n^F}(w|\Df) \beta_n^F\right]\\
\OU_w&=+\sum_{n=0}^\infty\left[ G_{\Delta_n^B}(w|\Df) \alpha_n^B+\partial_\Delta G_{\Delta_n^B}(w|\Df)\beta_n^B\right]\,.
\eea
Details of these and other formulae below will be given in the main body of the paper. The master functionals can be computed either from the above definitions, or more importantly and as we show, by solving a set of equations that make no reference to the functional bases.  

Section \ref{sec:dispersion} concerns the relation between the master functionals and the Polyakov bootstrap, as well as the derivation of new crossing-symmetric dispersion relations on the line. Concretely, Polyakov blocks can be computed from the master functionals above via the formulae
\bea
\mathcal P^F_\Delta(w|\Df)&:=\OL_w(\Delta|\Df)+G_{\Delta}(w|\Df)\,,\\
-\mathcal P^B_\Delta(w|\Df)&:=\OU_w(\Delta|\Df)-G_{\Delta}(w|\Df)\,.
\eea
This means that the sum rules associated to the master functionals can be reinterpreted as stating the validity of the Polyakov bootstrap, e.g.:
\bea
\sum_{\Delta} a_\Delta \OU_w(\Delta)=0 \qquad \Leftrightarrow \qquad \sum_{\Delta} a_\Delta \mathcal P^B_{\Delta}(w)=\sum_{\Delta} a_{\Delta} G_{\Delta}(w|\Df)=\mathcal G(w)
\eea
Using a representation of the functional action which makes manifest its relation to the double discontinuity leads to new expressions for Polyakov blocks and, via the Polyakov bootstrap, dispersion relations for CFT correlators:
\bea
\underline{\mathcal G}(w)&=+\int_0^1\ud z \, \ghl_w(z|\Df)\, \mbox{dDisc}_F\left[\underline{\mathcal G}(z)\right]\\
\overline{\mathcal G}(w)&=-\int_0^1\ud z \, \ghu_w(z|\Df)\, \mbox{dDisc}_B\left[\overline{\mathcal G}(z)\right]
\eea
where $\gl_w, \gu_w$ are kernels defining the master functionals. Strikingly, we will show they may be independently determined merely by assuming the existence of dispersion relations of the form above, without ever making any mention of functionals.

Section \ref{sec:appbounds} goes back to the original motivation of setting bounds on CFT correlators. Using the dispersion relations and the positivity properties of the master functionals we prove bounds on CFT correlators restricted to the line $z=\bar z$. These bounds are stronger than, but also imply the simple results:
\bea
\mathcal G(w)&\geq \mathcal G^F(w)=-1+\frac{1}{w^{2\Df}}+\frac{1}{(1-w)^{2\Df}}\,,\\
\mathcal G(w)&\leq \mathcal G^B(w)=+1+\frac{1}{w^{2\Df}}+\frac{1}{(1-w)^{2\Df}}\, \qquad \mbox{if} \quad \Delta_g\geq 2\Df
\eea
where $\Delta_g$ is the scaling dimension of the first non-identity operator appearing in the correlator\,. The lower (upper) bounds correspond to generalized free fermion (boson) correlators. Note that any upper bound on the correlator restricted to the line automatically implies a bound on the whole Euclidean section, since $\mathcal G(z,\bar z)\leq \mathcal G(\sqrt{z \bar z},\sqrt{z \bar z}).$\footnote{This follows from the interpretation of the correlator as the overlap between states in Hilbert space \cite{Pappadopulo:2012jk}.}

Section \ref{sec:regge} is devoted to a study of the CFT Regge limit, since this limit is accessible even on the line $z=\bar z$. We formulate the Regge limit in terms of power-law behaviour of the correlator, without making reference to Regge trajectories\,,
\bea
\, (1-z)^{2\Df} \left[\mathcal G(e^{2\pi i} z, z)-\mathcal G(z,z)\right]& \underset{z\to 1}\sim 
\mu \, (1-z)^{1-j_0}\,. \\
(1-z)^{2\Df} \mbox{dDisc}_s \mathcal G(z,z)
&\underset{z\to 1}{\sim} \nu (1-z)^{1-\tj}\,.
\eea
Exploiting the dispersion relations we derived, we go on to establish various bounds and relations between $\tj$  and $j_0\geq \tj$. The upshot is that (modulo subtleties which we discuss), $j_0$ is completely determined by the dispersion relation in terms of $\tj$ or the low dimension data of the CFT:
\bea
j_0=\mbox{max}\{\tj, 1-\Delta_g\}
\eea
In turn $\tj$ is determined by the behaviour of high dimension operators in the OPE via the relation
\bea
\overline{\gamma_n^2}\underset{n\to \infty}{\simeq}  \Delta_n^{-2(1-\tj)}\,, \qquad \overline{\gamma_n^2}:=\frac{1}{C_n} \sum_{|\Delta-\Delta_n|\leq C_n} \left(\frac{a_{\Delta}}{a_\Delta^{\mbox{\tiny free}}}\right) \frac{4 \sin^2\left[\frac{\pi(\Delta-2\Df)}2\right]}{\pi^2}
\eea
We also establish various bounds on $\tj$ along the way and show that $j_0<0$ implies constraints on the CFT spectrum.

In section \ref{sec:approxIsing} we use the Polyakov bootstrap to efficiently approximate the 3d Ising spin-correlator. This is possible since many of the low-lying operators in the correlator have dimensions close to that of a generalized free field. We find that a two block approximation approximates the correlator on the line $z=\bar z$ to about $0.1\%$. We go on to show that an even more spartan  but almost equally good  approximation is possible with a single (interacting) 1d Polyakov block 

We comment on the relation of this work to similar ones in higher dimensions in section \ref{sec:discussion} and conclude with perspectives on future work. This paper is complemented by several  appendices containing supplementary material and technical details.

\section{Review: functional bases for the 1d crossing equation}
\label{sec:review}
\subsection{Kinematics} In this paper we will be interested in studying CFT correlators of identical fields from the simplified point of view of their restriction to the line $z=\bar z$:
\bea
\langle \phi(x_1)\phi(x_2)\phi(x_3)\phi(x_4)\rangle:=\frac{\mathcal G(z,\bar z)}{x_{13}^{2\Df} x_{24}^{2\Df}}\,, \quad z \bar z=\frac{x_{12}^2 x_{34}^2}{x_{13}^2 x_{24}^2}\,,\quad (1-z)(1-\bar z)=\frac{x_{14}^2 x_{23}^2}{x_{13}^2 x_{24}^2} \,.
\eea

 Some care is needed regarding this restriction, since it implies the correlator becomes a piecewise analytic function of $z$ on the intervals $(-\infty,0)\cup(0,1)\cup(1,\infty)$. Accordingly we define the restriction to each subinterval to be $\mathcal G_-(z),\mathcal G_0(z), \mathcal G_+(z)$. The point is that by taking $z\neq \bar z$ we are allowed to analytically continue between branches, but we forfeit this possibility after imposing $z=\bar z$. These correlators satisfy the crossing relations:
\bea
\mathcal G_0(z)=\mathcal G_0(1-z)\,\qquad \mathcal G_-(z)=\eta \frac{\mathcal G_0(\mbox{$\frac{z}{z-1}$})}{(1-z)^{2\Df}}\,\qquad \mathcal G_+(z)=\eta\frac{\mathcal G_0(\mbox{$\frac 1z$})}{z^{2\Df}}\,.
\eea
where $\eta=+1 (-1)$ for bosonic (fermionic) $\phi$. We will focus on $\mathcal G_0(z)$, setting $\mathcal G\equiv \mathcal G_0$.

The function $\mathcal G_0(z)$ is actually an analytic function on $\mathbb C\backslash (-\infty,0]\cup [1,\infty)$ \cite{Pappadopulo:2012jk,Hogervorst2013}. Notice that $\mathcal G(z)$ arises from a higher dimensional correlator, complex $z$ necessarily corresponds to a Lorentzian configuration of operators. The continuation to complex $z$ of $\mathcal G_0(z)$ can be obtained from that of the full correlator $\mathcal G(z,\bar z)$ by first starting in the Euclidean section with $\bar z=z^*$ and continuing by taking $\bar z$ to $z$, crossing the real axis on the interval $(0,1)$. For instance, the Wightman correlator with pairs of operators (1,2) and (3,4) timelike separated can be obtained from  $\mathcal G_0(z+i\epsilon)$ with real $z<0$ and infinitesimal $\epsilon>0$.

The conformal block expansion of $\mathcal G(z)$ can be written as follows:
\bea
\mathcal G(z)=\sum_{\Delta} a_\Delta G_{\Delta}(z|\Df)\,,\qquad z\in \mathbb C\backslash (-\infty,0]\cup[1,\infty)\,.
\eea
where $a_{\Delta}$ are squared OPE coefficients $a_{\Delta}:=\lambda_{\phi \phi \mathcal O_{\Delta}}^2$ and the $SL(2,\mathbb R)$ conformal blocks are given by \cite{Dolan:2003hv}:
\bea
G_{\Delta}(z|\Df)=z^{\Delta-2\Df}\, _2F_1(\Delta,\Delta,2\Delta,z)
\eea
In general, thinking of $\mathcal G$ as a line restriction of a higher-D correlator, these capture only contributions from an $SL(2,\mathbb R)$ subgroup of the full conformal group. In particular, higher dimensional conformal blocks, including ones with non-zero spin $\ell$, correspond to infinite sums of the ones given above with positive coefficients \cite{Hogervorst:2016hal}. Accordingly, the corresponding OPE coefficients $a_{\Delta}$ above might not all be independent. In this paper we will mostly ignore this possible higher dimensional origin and consider the most general case possible, which includes one-dimensional theories. The crossing equation for $\mathcal G$ becomes:
\bea
\sum_{\Delta\geq 0} a_{\Delta} F_{\Delta}(z|\Df)=0, \qquad F_{\Delta}(z|\Df)=G_{\Delta}(z|\Df)-G_{\Delta}(1-z|\Df)\,. \label{eq:crossing}
\eea
keeping in mind the constraint $a_{\Delta}\geq 0$ for unitarity theories.

As we will see, an important role is played by two special correlators which describe the fundamental field four-point function of generalized free fields. These come in bosonic and fermionic varieties, with corresponding correlators:
\bea
\mathcal G^B(z)&=\frac{1}{z^{2\Df}}+\frac{1}{(1-z)^{2\Df}}+1=\frac{1}{z^{2\Df}}+\sum_{n=0}^\infty a_{\Delta_n^B}^{\mbox{\tiny free}} G_{\Delta_n^B}(z|\Df)\,,& \Delta_n^B&=2\Df+2n\\
\mathcal G^F(z)&=\frac{1}{z^{2\Df}}+\frac{1}{(1-z)^{2\Df}}-1=\frac{1}{z^{2\Df}}+\sum_{n=0}^\infty a_{\Delta_n^F}^{\mbox{\tiny free}} G_{\Delta_n^F}(z|\Df)\,,& \Delta_n^F&=1+2\Df+2n
\eea
where
\bea
a_{\Delta}^{\mbox{\tiny free}}=\frac{2 \Gamma(\Delta)^2}{\Gamma(2\Delta-1)}\, \frac{\Gamma(\Delta+2\Df-1)}{\Gamma(2\Df)^2\Gamma(\Delta-2\Df+1)} \label{eq:ope1d}\,,
\eea
is what we call the ``free'' OPE density.
\subsection{Functional bases} A general method for studying the consequences of the crossing equation is to act with suitable linear functionals on the crossing equation \reef{eq:crossing} \cite{Mazac:2016qev,Mazac:2018,Mazac2019a}. A general class of functionals is described by the ansatz:
\bea
\omega\left[\mathcal F\right]=\int_1^\infty \frac{\ud z}{\pi}\, h(z) \mathcal I_z \mathcal F(z)\,,\label{eq:funch}
\eea
where for $\mathcal F(z^*)=\mathcal F^*(z)$ we define
\bea
\mathcal I_z \mathcal F(z)=\lim_{\epsilon\to 0^+} \frac{\mathcal F(z+i\epsilon)-\mathcal F(z-i\epsilon)}{2i}\,,\qquad \mathcal R_z \mathcal F(z)=\lim_{\epsilon\to 0^+} \frac{\mathcal F(z+i\epsilon)+\mathcal F(z-i\epsilon)}{2}\,.
\eea
The function $\mathcal F(z)=-\mathcal F(1-z)$ should be analytic in $\mathbb C\backslash(-\infty,0]\cup[1,\infty)$. If we take $h(z)$ analytic away from the real line and suitably bounded, we can rewrite the above as
\bea
\omega[\mathcal F]=\frac 1{2} \int_{\frac 12}^{\frac 12+i\infty} \ud z f(z) \mathcal F(z)+\int_{\frac 12}^1 \ud z\, g(z) \mathcal F(z)\label{eq:funcfg}
\eea
where 
\bea
f(z)=\frac{h(z)-h(1-z)}{i\pi}\,,\quad \mbox{Im}\, z>0\,,\qquad g(z)=-\frac{\mathcal I_z h(z)}\pi, \qquad z\in(0,1) \label{eq:fgfromh}
\eea
which implies the so-called gluing condition:
\bea
\mathcal R_z f(z)=-g(z)-g(1-z)\,.\label{eq:gluing}
\eea
As a trivial example, the evaluation functional $\mathcal E_w$ defined by
\bea
\mathcal E_w\left[\mathcal F\right]:=\mathcal F(w)
\eea
can be written as above by choosing
\bea
\mathcal E_w:\qquad &h(z)=\frac{1-2w}{(z-w)(1-z-w)}\,,\label{eq:idfunc1}
\eea
or equivalently
\bea
g(z)=\delta(z-w)-\delta(1-z-w)\,,\qquad f(z)=0\,.\label{eq:idfunc2}
\eea
Far less trivial sets of functionals were constructed in \cite{Mazac2019a}. These sets are in a sense dual to the bosonic and fermionic free correlators mentioned above:
\vspace{0.3cm}
\bea
\mbox{Fermionic basis:}\qquad &\mathcal S_F=\{\alpha_n^F, \beta_n^F,\quad n\in \mathbb Z_{\geq 0}\}\\
\mbox{Bosonic basis:}\qquad &\mathcal S_B=\{\alpha_n^B, \beta_n^B,
\quad n\in \mathbb Z_{\geq 0}\}\,.\vspace{0.3cm}
\eea
The functionals in these bases satisfy the duality conditions:
\bea
\alpha_n^F(\Delta_m^F)&=\delta_{n,m},& \partial_{\Delta}\alpha_n^F(\Delta_m^F)&=0,\\
\beta_n^F(\Delta_m^F)&=0,& \partial_{\Delta}\beta_n^F(\Delta_m^F)&=\delta_{n,m}\,,\label{eq:dualityfermion}
\eea
and
\bea
\alpha_n^B(\Delta_m^B)&=\delta_{n,m},& \partial_{\Delta}\alpha_n^B(\Delta_m^F)&=-c_{n}\delta_{m0},\\
\beta_n^F(\Delta_m^F)&=0,& \partial_{\Delta}\beta_n^F(\Delta_m^F)&=\delta_{n,m}-d_n \delta_{m0}\,,\label{eq:dualityboson}
\eea
for some constants $c_n, d_n$ which can be determined explicitly. The construction of the kernels $f,g$ for these functionals was given in detail in \cite{Mazac:2018,Mazac2019a}.

All these functionals are {\em crossing-compatible}, meaning that their action commutes with the infinite sum of states in the crossing equation \cite{Rychkov:2017tpc}. In this way, any such functional leads to a necessary condition on the CFT data, expressed by a sum rule:
\bea
\omega\left[\sum_{\Delta} a_{\Delta} F_{\Delta}\right]=\sum_{\Delta} a_{\Delta} \omega(\Delta)=0\,.\label{eq:compatib}
\eea
Technically the crossing-compatible condition can be 
satisfied by demanding that the kernel $f(z)$ decays at large $z$ as $z^{-1-\epsilon}$ for some positive $\epsilon$.

The key point is that these sets of functionals fully capture the constraints of crossing symmetry, in the sense that:
\bea
&\sum_{\Delta} a_{\Delta} F_{\Delta}(z)=0& \qquad &\mbox{for all} \quad z\in \mathbb C\backslash (-\infty,0)\cup (1,\infty)\\
\Leftrightarrow \qquad &\sum_{\Delta} a_{\Delta} \omega(\Delta)=0& \qquad &\mbox{for all}\quad \omega\in \mathcal S_F\\
\Leftrightarrow \qquad &\sum_{\Delta} a_{\Delta} \omega(\Delta)=0& \qquad &\mbox{for all}\quad \omega\in \mathcal S_B\,. \label{eq:completeness}
\eea
Notice that while one of the directions of implication is trivial using \reef{eq:compatib}, the other one is not, as it establishes ``completeness'' of the functionals in the bases $\mathcal S_F$ and $\mathcal S_B$.
We will comment on the meaning of completeness in subsection \ref{sec:polycomplete}.

Before we continue, a small remark on notation. Given a functional $\omega$ we define
\bea
\omega\left[F_{\Delta}(\bullet|\Df)\right]\equiv \omega(\Delta|\Df)
\eea
We will sometimes drop the dependence on $\Df$, i.e. $\omega(\Delta|\Df)\to \omega(\Delta)$ to unclutter the notation.

\section{Master Functionals}
\label{sec:master}
In this section we introduce the concept of master functionals. We begin by motivating them by posing the problem of setting bounds on values of CFT correlators. We show that the functionals are defined by certain equations satisfied by the associated kernels, and that these equations can be solved efficiently numerically.

\subsection{Motivation: correlator bounds}
\label{sec:extfuncs}
Suppose we would like to find general bounds on the allowed values of CFT correlators. This is an interesting problem which will be studied more completely in upcoming work \cite{toappearZechuan}. Here we focus on the simplest version of this problem which is to choose a point $w \in (0,1)$ and ask for bounds such that:
\bea
\mathcal G_{\mbox{\tiny min}}(w) \leq \mathcal G(w)\leq \mathcal G_{\mbox{\tiny max}}(w)\,,
\eea
for any unitary CFT correlator $\mathcal G(w)$.  A subtlety is that it is not necessarily the case that the upper or lower bounds, as functions of $w$, are given by actual CFT correlators. At most, all we can say is that for some choice of $w$, there exist some CFT correlator which will saturate the bounds for that given $w$. As we'll see shortly, in this work this complication relevant.

We can have an idea of what to expect by using the OPE. For small $w$ we have
\bea
w^{2\Df} \mathcal G(w)\sim 1+a_{\Delta_g} w^{\Delta_g}+\ldots
\eea
where $\Delta_g$ is the dimension of the first non-identity state in the OPE. For sufficiently small $w$ it is clear that minimizing the correlator should be essentially the same as maximizing the gap $\Delta_g$.\footnote{This can be used to avoid costly computational searches in solving the gap maximization problem, as will be explored in \cite{toappearZechuan}.} But this gap cannot be arbitrarily large: there is a universal upper bound on the gap equal to $\Delta_g=1+2\Df$ saturated by the generalized free fermion solution discussed in the previous section \cite{Mazac:2016qev,Mazac:2018}. Hence we tentatively expect:
\bea
\mathcal G(w)\geq \mathcal G^F(w)\, \qquad \mbox{for all}\quad w \in (0,1)\,.
\eea
This will turn out to be correct.

The same logic tells us that maximizing the correlator will not give us anything interesting, at least in $d=1$, since in this case $\Delta_g\geq 0$: maximization will just push down $\Delta_g$ towards zero and at some point the corresponding OPE coefficient will become unbounded. We can remedy this by asking what is the maximal allowed value of a CFT correlator for a fixed value of the gap $\Delta_g$. In this case we expect that the maximal possible value of the correlator will be achieved, at least for small $w$, by maximizing the OPE coefficient $a_{\Delta_g}$. Depending on the value of the gap there is indeed an upper bound on this coefficient. A particularly interesting case corresponds to setting $\Delta_g=2\Df$, since in this case the bound exists and is saturated by the generalized free boson solution. Hence we expect
\bea
\mathcal G(w)\leq \mathcal G^B(w) \, \qquad \mbox{for all}\quad w \in (0,1)\,, \qquad \mbox{if} \quad \Delta_g\geq 2\Df\,.
\eea
Again, we will see that this expectation is correct. For other choices of gap we should have that $\mathcal G(w)$ is bounded from above by the correlator saturating the OPE maximization problem, which in general is described by a non-trivial (interacting) solution to crossing. This solution can be efficiently determined by numerical methods \cite{Paulos:2019fkw}. For simplicity we will not explore this possibility further in this work.

Since we now have candidate optimal bounds on the correlator we can set about constructing functionals which will prove them.

\paragraph{Correlator minimization} Let us discuss first the minimization problem.  Suppose we are given a functional $\Omega_w$ satisfying
\bea
\Omega_{w}(\Delta)\geq -G_{\Delta}(w|\Df)\,,\qquad \mbox{for all}\quad \Delta>0\,. \label{eq:poslower}
\eea
Acting with such a functional on the crossing equation gives
\bea
0=\Omega_w(0)+\sum_{\Delta>0}a_{\Delta} \Omega_w(\Delta)\Rightarrow \sum_{\Delta>0} a_{\Delta} G_{\Delta}(w|\Df)\leq \Omega_w(0)
\eea
and so we have
\bea
\mathcal G(w)\geq G_{0}(w|\Df)+\Omega_w(0)\,.
\eea
In general given some basis of functionals we can look for linear combinations satisfying the positivity conditions above and obtain the one which minimizes $\Omega_w(0)$. 

However, in our case we have a candidate optimal, exact, bound. Let us suppose that the minimizing correlator is indeed that of a generalized free fermion. The only way in which this is possible is if the constraints on the corresponding optimal functional $\OL_w$ are saturated whenever the scaling dimension $\Delta$ happens to lie in the spectrum of the GFF correlator. 
\bea
\OL_{w}(\Delta_n^F)=-G_{\Delta_n^F}(w|\Df),\qquad \partial_\Delta \OL_{w}(\Delta_n^F)=-\partial_\Delta G_{\Delta_n^F}(w|\Df)\,, \qquad n=0,1,\ldots \label{eq:posfermion}
\eea
With the first set of conditions it is easy to show (by acting with the functional on the crossing equation) that $\OL_w(0)=\mathcal G^F(w)-G_0(w|\Df)$ as we want. As for the derivative conditions, these are necessary to ensure that the positivity constraints on the functional hold in open neighbourhoods around $\Delta=\Delta_n^F$. 
Using the basis of functionals described in the previous subsection it is a simple matter to construct the unique functional which satisfies these conditions:
\begin{eqnarray}
\boxed{\OL_w=-\sum_{n=0}^{+\infty} \left[G_{\Delta_n^F}(w|\Df) \alpha_n^F+\partial_\Delta G_{\Delta_n^F}(w|\Df) \beta_n^F\right]}\label{eq:sumexprl}
\end{eqnarray}
We say that $\Omega^F_w$ is the {\em master functional} for the fermionic functional basis. 
This is our candidate extremal functional 
for the correlator minimization problem. At this point it is only a candidate, since we need to show that the positivity conditions \reef{eq:posfermion} also hold away from the special values $\Delta=\Delta_n^F$. We will postpone this however to section \ref{sec:appbounds} and turn now to the analogous problem for correlator maximization.

\paragraph{Correlator maximization}
The logic here is similar to the one for the minimization problem, all that is required is to flip some signs. An upper bound on a correlator whose first non-identity state has $\Delta\geq \Delta_g=2\Df$ is determined by constructing a functional $\Omega_w$ satisfying the positivity conditions
\bea
\Omega_w(\Delta)\geq G_{\Delta}(w|\Df)\,, \qquad \mbox{for}\quad \Delta\geq \Delta_g=2\Df\,.
\eea
Applying such a functional to the crossing equation is easily seen to lead to the bound:
\bea
\mathcal G(w)\leq G_0(w|\Df)-\Omega_w(0)
\eea
Again, a general bound can be found by constructing a functional from some basis such that the positivity conditions above hold, and then maximizing $\Omega_w(0)$

Following the same logic as for the minimization problem, we expect that the positivity conditions will be saturated on the GFB spectrum:
\bea
\OU_w(\Delta_n^B)&=G_{\Delta_n^B}(w|\Df)\,,&\quad n&=0,1,\ldots\\
\partial_{\Delta}\OU_w(\Delta_n^B)&=\partial_{\Delta} G_{\Delta_n^B}(w|\Df)\,,& n&=1,2,\ldots \label{eq:posboson}
\eea
What is new now is that for the set of derivative conditions we do not expect the $n=0$ equation to hold, since this is the statement that the maximization problem is sensitive to the choice of gap. Using the bosonic basis of functionals we can again easily write the unique functional satisfying these constraints. The {\em master functional} for the bosonic functional basis is:
\begin{eqnarray}
\boxed{
\OU_w=\sum_{n=0}^\infty\left[ G_{\Delta_n^B}(w|\Df) \alpha_n^B+\partial_\Delta G_{\Delta_n^B}(w|\Df)\beta_n^B\right]}\label{eq:sumexpru}
\end{eqnarray}
We remind the reader that $\beta_0^B\equiv 0$, and this is consistent with the absence of the $n=0$ equation in the derivative constraints written above.

Let us conclude with a quick recap. We have used the functional bases reviewed in the previous subsection to define master functionals $\Omega^{B,F}_w$. Subject to (as yet) unchecked positivity conditions being satisfied, the master functionals are extremal functionals for correlator minimization and maximization problems establishing the rigorous bounds:
\bea
\mathcal G^F(w)\leq \mathcal G(w)\leq \mathcal G^B(w)
\eea
where for the upper bound we have assumed the gap $\Delta_g=2\Df$ in the spectrum of $\mathcal G$. We will show the right positivity conditions hold in section \ref{sec:appbounds}.

\subsection{Equations and boundary conditions}

The goal of this section is to provide a direct  definition of the master functionals that does not rely on their representation in terms of the 1d functional bases.

\paragraph{Master functional equations}
Our starting point is to assume a representation of the form presented in \reef{eq:funcfg} for the functional action:
\bea
\Omega_w^{B,F}(\Delta|\Df)=\frac 1{2} \int_{\frac 12}^{\frac 12+i\infty} \ud z \ful_w(z) F_\Delta(z|\Df)+\int_{\frac 12}^1 \ud z\, \gul_w(z)  F_\Delta(z|\Df)\label{eq:funcactionfg}
\eea
We will choose the kernels appearing in above expression in such a way that the extremality conditions \reef{eq:posfermion} and \reef{eq:posboson} are be satisfied. Based on previous experiences with the bosonic and fermionic functional bases this is not so hard to do \cite{Mazac2019a}. We set $\ful_w(z)$ holomorphic away from the interval $z
\in (0,1)$ and choose:
\bea
\gl_w(z)&=-\delta(z-w)+\ghl_w(z)\,,&\qquad \fl_w(z)&=-(1-z)^{2\Df-2} \ghl_w\left(\mbox{$\frac 1{1-z}$}\right)\\
\gu_w(z)&=+\delta(z-w)+\ghu_w(z)\,,&\qquad \fu_w(z)&=+(1-z)^{2\Df-2} \ghu_w\left(\mbox{$\frac 1{1-z}$}\right) \label{eq:relfg}
\eea
Recall however that we are not completely free in choosing the $f,g$ kernels, since the gluing condition \reef{eq:gluing} must be respected. These become the equations:
\begin{subequations}
\begin{empheq}[box=\fbox]{align}
\mathcal R_z \fl_w(z)-(1-z)^{2\Df-2}\fl_w(\mbox{$\frac 1{1-z}$})-z^{2\Df-2}\fl_w(\mbox{$\frac 1{z}$})&=+\delta(z-w)+\delta(1-z-w)\\
\mathcal R_z \fu_w(z)+(1-z)^{2\Df-2}f_w^u(\mbox{$\frac 1{1-z}$})+z^{2\Df-2}f_w^u(\mbox{$\frac 1{z}$})&=-\delta(z-w)-\delta(1-z-w) 
\end{empheq}\label{eq:fundfreeeq}
\end{subequations}
These are the equations that we must try to solve, subject to:
\bea
\boxed{
\ful_w(z)\underset{z\to\infty}{=}O(z^{-2})\,,\qquad \fl_w(z)\underset{z\to 0^-}=O(\log(-z))\,,\qquad \fu_w(z)\underset{z\to 0^-}=O(z^{-1})}\label{eq:boundcond}
\eea
The first boundary condition, which concerns the behaviour of the kernel at infinity, is the usual condition necessary to ensure that the functional is crossing-compatible. We will discuss the other conditions below. The claim is that there is a unique solution to these equations with this choice of boundary conditions, and that it defines functionals satisfying conditions \reef{eq:posfermion} or \reef{eq:posboson} accordingly. In appendix \ref{app:funcsol} we explain how the kernels may be computed analytically by solving these equations for special values of $\Df$, or in a perturbative expansion around large $z$. More generally, in section \ref{sec:fredholm} below we will see that the kernels can be computed numerically in full generality easily and efficiently by solving a standard integral equation.

Why did we choose to relate $f,g$ via \reef{eq:relfg}? Well, the point is that thanks to this choice we can show (by a contour deformation of the integral in \reef{eq:funcactionfg}) that:
\bea
\OL_w(\Delta)+G_{\Delta}(w|\Df)&=
\cos^2\left[\frac{\pi}2(\Delta-2\Df)\right] \int_0^1 \ud z \ghl_w(z) G_{\Delta}(z|\Df)\,,&\qquad \Delta&>2\Df-1\\
\OU_w(\Delta)-G_{\Delta}(w|\Df)&=\sin^2\left[\frac{\pi}2(\Delta-2\Df)\right] \int_0^1 \ud z \ghu_w(z) G_{\Delta}(z|\Df)\,,& \qquad \Delta&>2\Df\,.\label{eq:funcactions}
\eea
In this way the functional actions do indeed satisfy the necessary conditions formulated in \reef{eq:posfermion} and \reef{eq:posboson}.  Whether the positivity conditions hold for general $\Delta$ away from $\Delta_n^{B,F}$ depends on the details of the kernels. In particular we see that one way to realize the positivity conditions (at least for the regions denoted above), would be for the $\gul_w$ kernels to be positive when both $w,z$ lie inside the interval $(0,1)$. This is indeed what we find in all cases where we have been able to construct the kernels (cf. figure \ref{fig:kernels}). 

\paragraph{Boundary conditions} Let us now discuss the boundary condition at $z=0$, and in particular how this relates to the regions of validity of the above expressions. The boundary conditions on $\ful_w$ translate into the behaviour of $\ghul_w$ near $z=0$. Since in this region we have $G_{\Delta}(z|\Df)\sim z^{\Delta-2\Df}$ the integrals above may develop singularities arising from the small-$z$ integration region for sufficiently small $\Delta$. This does not mean that the functional action diverges there. Indeed it is easy to see that the original definition \reef{eq:funcactionfg} together with the choice of boundary conditions imply that the functional action is certainly finite for $\Delta\geq 0$. Instead it is the contour deformation argument which fails. In equations \reef{eq:funcactions} we have provided explicit regions of validity of the representation which follow from our choice of boundary conditions. But how do we know which ones to choose and more importantly whether they lead to a unique solution? The reason is that equations \reef{eq:fundfreeeq} have infinite sets of ``homogeneous'' solutions, i.e. choices of kernels $f$ which satisfy these equations up to the delta function terms. These solutions are nothing but the ordinary basis functionals $f_{\alpha_n}, f_{\beta_n}$. Indeed, for those solutions we have not only the correct fall off at $z=\infty$ but also:
\bea
g_{\omega^F}(z)=-(1-z)^{2\Df-2} f_{\omega^F}(\mbox{$\frac{1}{1-z}$})\,,\qquad g_{\omega^B}(z)=(1-z)^{2\Df-2} f_{\omega^B}(\mbox{$\frac{1}{1-z}$})
\eea
and the asymptotic behaviours
\bea
g_{\alpha_n^F}(z)\underset{z\to 0}{\propto} -z^{-2-2n} \log(z)\,,\qquad
g_{\beta_n^F}(z)\underset{z\to 0}{\propto} z^{-2-2n}\\
g_{\alpha_n^B}(z)\underset{z\to 0}{\propto} -z^{-1-2n} \log(z)\,,\qquad
g_{\beta_n^B}(z)\underset{z\to 0}{\propto} z^{-1-2n}
\eea
Our boundary conditions were chosen such that homogenous terms must all be fixed to ensure that
\bea
\gl_w(z)\underset{z\to 0}{=} O(\log(z))\,,\qquad \gu_w(z) \underset{z\to 0}{=} O(z^{-1})\,.
\,,
\eea
Notice that naively one may think that it would only be possible to set $\gl_w(z)$ to be $O(z)$ but this cannot be, otherwise the functional action would diverge when taking $\Delta\to 2\Df$. Incidentally this is the same reason why the power-law behaviour of functional basis elements changes in steps of two.

An important property of the equations for the master functional kernels is that they have a symmetry under the exchange $w\leftrightarrow 1-w$. This translates into the same symmetry for the functional kernels and therefore any bounds that we derive from them will also be manifestly crossing symmetric. Focusing on the minimizing functional we have
\bea
\fl_w(z)-\fl_{1-w}(z)&=\ghl_w(z)-\ghl_{1-w}(z)=0\\
\Rightarrow \gl_{w}(z)-\gl_{1-w}(z)&=-\delta(z-w)+\delta(1-z-w)
\eea
Recalling \reef{eq:idfunc1} and \reef{eq:idfunc2} this tells us that
\bea
\OL_w-\OL_{1-w}=-\mathcal E_w \qquad \Rightarrow \qquad \OL_w(\Delta|\Df)-\OL_{1-w}(\Delta|\Df)=-F_{\Delta}(w|\Df).
\eea
As a consequence the lower bound on the correlator, which is determined by $\OL_w(0)+G_0(w|\Df)$, is crossing symmetric as it should be. Similar statements apply to the $\OU_w$ functional.

Let us conclude by pointing out that the symmetry under $w\leftrightarrow 1-w$ is quite non-trivial from the perspective of expressions \reef{eq:sumexprl}, \reef{eq:sumexpru} for the master functionals written in terms of the 1d functional bases. Symmetry implies that $\fl_w(z)$ must act as a generator of crossing symmetric functions in $w$ made up entirely of conformal blocks and their derivatives with scaling dimensions $\Delta_n^F$. Such functions are spanned by certain contact interaction in AdS space. We examine this more closely in appendix \ref{app:funcsol}.

\subsection{General solution}
\label{sec:fredholm}
\begin{figure}%
\begin{center}
\begin{tabular}{cc}
\includegraphics[width=8.1 cm]{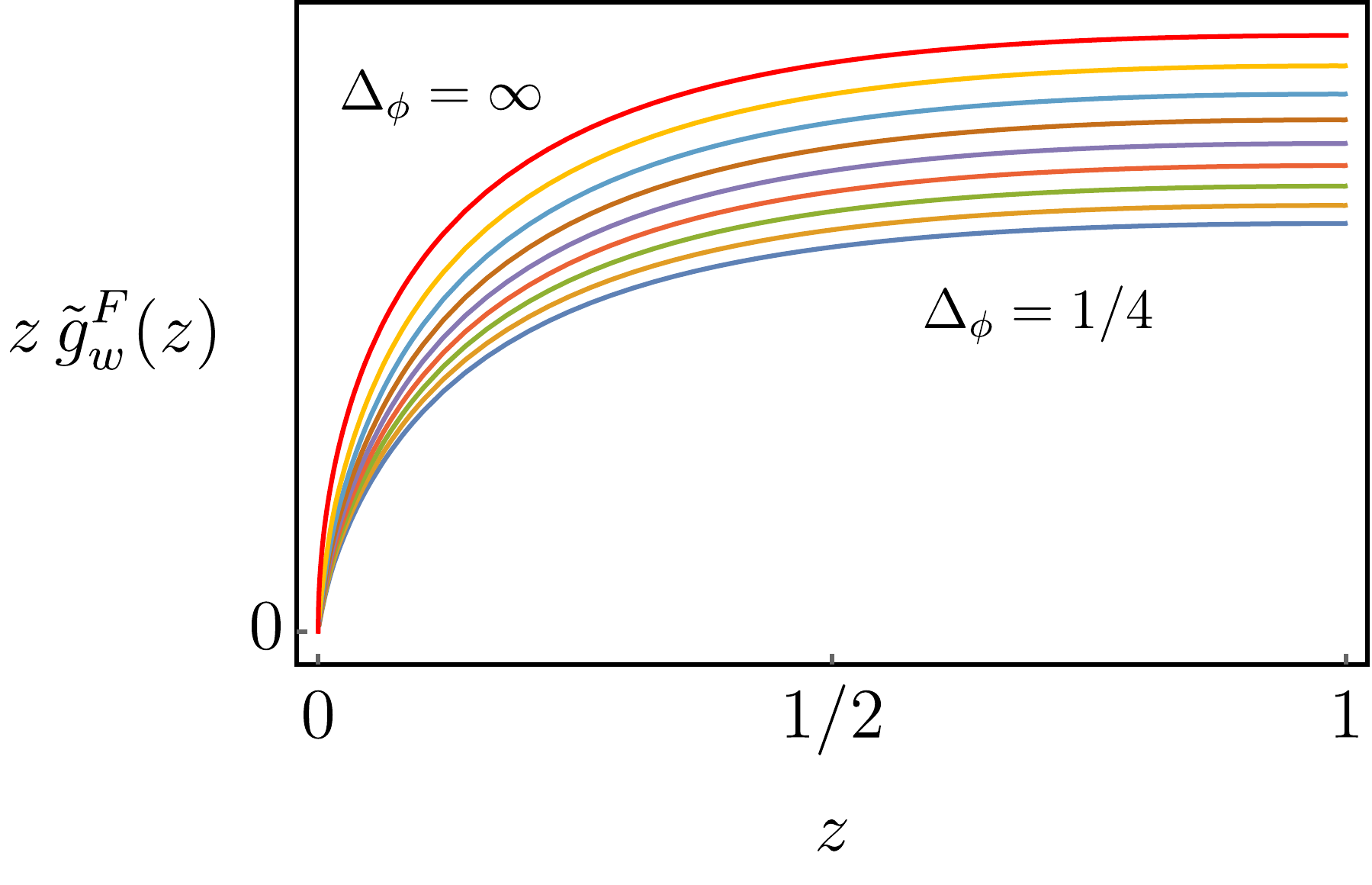}%
&
\includegraphics[width=8 cm]{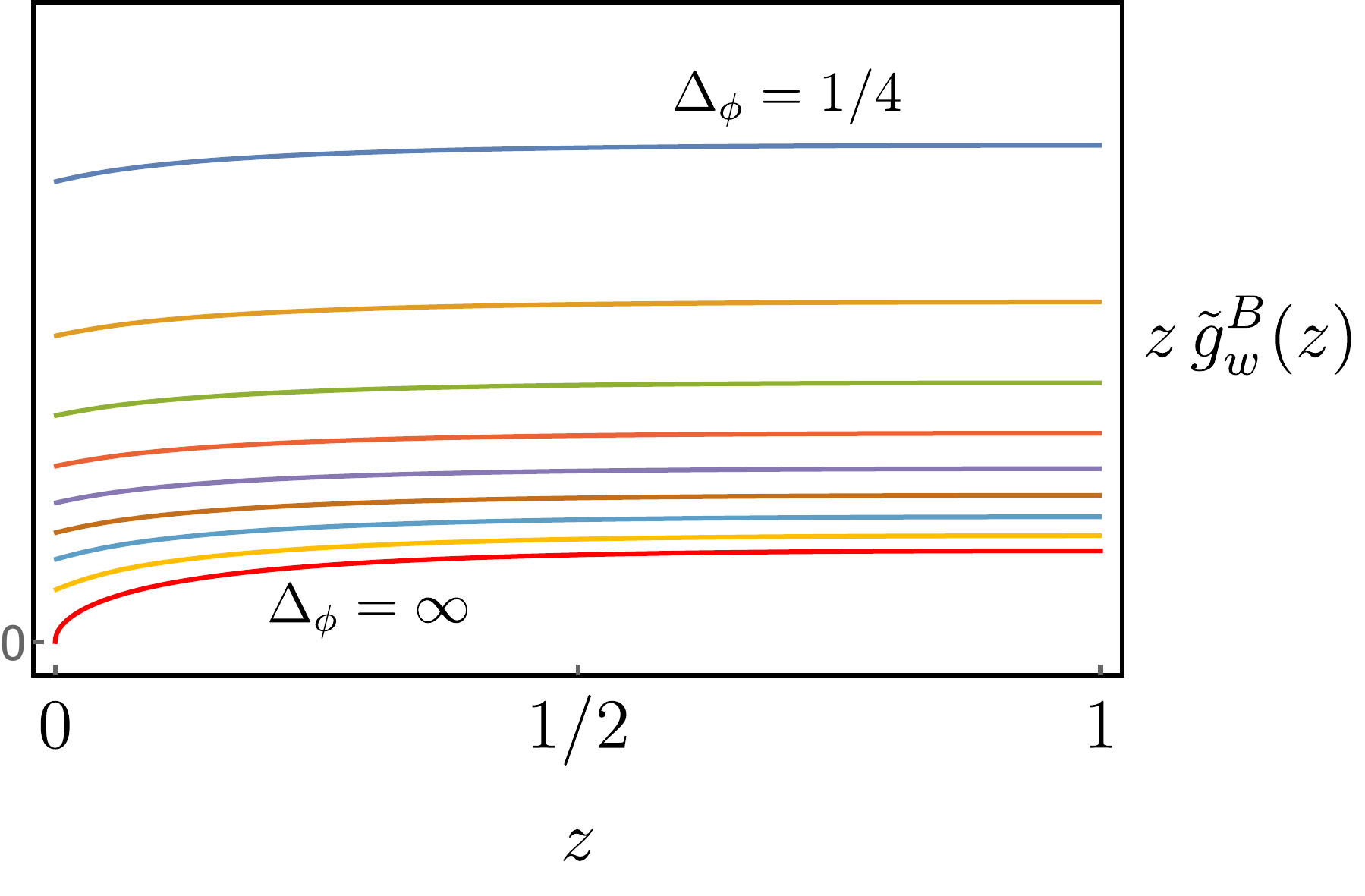}%
\end{tabular}
\caption{Functional kernels for values of $\Df$ ranging from $\Df\sim 0.25$ to $\infty$, evaluated for $w=1/3$. For clarity we plot the kernels in terms of $\tilde g^{B,F}_w(z)\equiv (1-z)^{-2\Df}\ghul_w(z)$. For special values of $\Df$ the numerical curves above match with analytic expresions. The kernels are positive for all $z\in(0,1)$. The rough shape of the curves does not change with $w$}%
\label{fig:kernels}%
\end{center}
\end{figure}

We will now show that the $f,g$ kernels satisfy integral equations which not only define them implicitly, but allow us to compute them numerically. We begin by setting:\footnote{\label{fn:zz}In this formula and elsewhere in this paper, when we write $\sqrt{z(z-1)}$ we really mean the function analytic in $\mathbb C\backslash [0,1]$ given by
\bea
\sqrt{z(z-1)}:=\left\{\begin{array}{cc}
-\sqrt{-z}\sqrt{1-z}\,, & z<0\\
\sqrt{z}\sqrt{z-1}\,, & \mbox{elsewhere}
\end{array}
\right.
\eea
}
\bea
\hat f_w(z):=\sqrt{z(z-1)} \ful_w(z)\,.
\eea
Note that $\hat f_w(z)=-\hat f_w(1-z)$  and by assumption it is analytic everywhere in $\mathbb C\backslash [0,1]$. Using Cauchy's formula we find:
\bea
\hat f_w(z)=\oint_{z} \frac{\ud z'}{2\pi i} \frac{\hat f_w(z)}{z'-z}=\frac{2}{\pi} \int_0^{1} \ud z' \frac{\frac 12-z}{(z'-z)(1-z'-z)}\, \mathcal I_{z'} \hat f_w(z')\,.
\eea
In the first equality the contour is a sufficiently small clockwise circle around $z$. To obtain the last formula we blow up the contour to wrap the discontinuity of $\hat f_w$ in the interval $[0,1]$ and used its antisymmetry under $z\leftrightarrow 1-z$. The boundary conditions \reef{eq:boundcond} imply that arcs at infinity can be dropped, and that wrapping contours around $0$ and $1$ is indeed allowed. We can now use \reef{eq:fundfreeeq} to obtain the discontinuity:
\begin{multline}
\pm \mathcal I_z \hat f_w(z)-(1-z)^{2\Df-\frac 12} \hat f_w(\mbox{$\frac 1{1-z}$})-z^{2\Df-\frac 12} \hat f_w(\mbox{$\frac 1{z}$})=\\=\sqrt{z(1-z)}\left[\delta(w-z)+\delta(1-w-z)\right]\,,
\end{multline}
with the positive sign corresponding to the $\fl_w$ case and the negative to $\fu_w$. Using this result we find
\begin{equation}
\label{eq:dispf}
\boxed{
\pm \ful_w(z)=\frac{2}{\pi}\sqrt{\frac{w(1-w)}{z(z-1)}} \frac{\frac 12-z}{(w-z)(1-w-z)}+\int_1^{\infty} \ud z' K_\Df(z,z') \ful_w(z')\,,
}
\end{equation}
with
\bea
K_\Df(z,z'):=\frac{2}{\pi} \sqrt{\frac{z'(z'-1)}{z(z-1)}}\, \frac{\frac 12-z}{(1-\frac 1{z'}-z)(\frac{1}{z'}-z)}\, (z')^{-2\Df-\frac 32}\,.
\eea
This integral equation is a Fredholm equation of the second kind depending on parameters $w$ and $\Df$. The boundary conditions \reef{eq:boundcond} are built in automatically to the equation. One can check that the equation is satisfied by the analytic solutions constructed in appendix \ref{app:funcsol} for special values of $\Df$. Furthermore the large $\Df$ solution is also easily extracted from this equation: it is simply the inhomogeneous term on the righthand side, as the contribution of the kernel $K_{\Df}$ becomes exponentially suppressed in this limit. For general values of $w, \Df$ the equation above can be readily solved numerically. Once we have solved it for $\ful_w(z)$ in the region $z>1$ it then provides us with the analytic continuation for any complex $z$. In figure \ref{fig:kernels} we show some of the functional kernels obtained solving the Fredholm equations by discretization.

There is a subtlety which we have omitted, which is that for the bosonic case the integral equation actually admits a homogeneous solution. This solution is the product of the $\alpha_0^B$ functional by a crossing symmetric function of $w$ (namely the contact term in $AdS_2$, see \ref{app:contacts}). This is allowed because the integral equation allows in general for a behaviour $f^B(z)\underset{z\to 1}\sim (1-z)^{-1}[a+b \log(z-1)]$, where we would like for $b$ to vanish according to our choice of boundary conditions. This ambiguity is not a problem when solving the Fredholm equation numerically, since such a solution involves a discretization that implicitly fixes the ambiguity. Since the $\alpha_0^B$ functional is known analytically for any $\Df$ \cite{Mazac:2018} we can subtract it out to obtain our desired solution. This can be done for instance by imposing the correct value of the master functional on the identity, which is the procedure we have implemented to obtain the kernels in figure \ref{fig:kernels}.

\subsection{Subtleties in the master functional definitions}
\label{sec:subtle}
In this section we mention a few subtleties regarding the definitions of the master functionals. The points made here are not crucial for the rest of this work and may be skipped on a first reading. For definiteness we will focus on $\OL_w$. 

We began by defining this functional in terms of the 1d functional basis expansion. This definition tells us how to determine the action of $\OL_w$ in terms of those of the basis elements:
\bea
\OL_w[\mathcal F]:=-\sum_{n=0}^{\infty}\left( G_{\Delta_n^F}(w|\Df) \alpha_n^F[\mathcal F]+
 \partial_\Delta G_{\Delta_n^F}(w|\Df) \beta_n^F[\mathcal F]\right)\,. \label{eq:def1}
\eea
We know that $\alpha_n^F,\beta_n^F$ admit a representation of the form \reef{eq:funch} in terms of kernels $h_{\alpha_n^F}, h_{\beta_n^F}$ and therefore
\bea
\OL_w[\mathcal F]=\sum_{n=0}^{\infty}\int_1^{\infty}\frac{\ud z}{\pi}\left[ G_{\Delta_n^F}(w|\Df) h_{\alpha_n^F}(z)+
 \partial_\Delta G_{\Delta_n^F}(w|\Df) h_{\beta_n^F}(z)\right]\mathcal I_z \mathcal F(z)\,. \label{eq:actionkernels}
\eea
The first question concerns whether, starting from the above, $\OL_w$ itself admits a similar representation:
\bea
\OL_w[\mathcal F]\overset{?}{=}\int \frac{\ud z}{\pi} h_w^F(z) \mathcal I_z \mathcal F(z)\,. \label{eq:rep1}
\eea
This does not follow immediately from \reef{eq:def1}. Technically we would need to exchange the order of summation and integration in \reef{eq:actionkernels} but whether this can be done depends on the set of test functions $\mathcal F$.\footnote{Note also that we cannot even show that the sums above converge since we have insufficient knowledge of the basis kernels for general $\Df$.} 
Choosing $\mathcal F(z)=1/(z-z_0)-1/(1-z-z_0)$ as our test functions it is easy to see that if that is the case then we would necessarily have
\bea
h_w^F(z)&=-\sum_{n=0}^{\infty}\left[ G_{\Delta_n^F}(w|\Df) h_{\alpha_n^F}(z)+\partial_{\Delta}G_{\Delta_n^F}(w|\Df) h_{\beta_n^F}(z)\right]\,.
\eea
Our second definition of $\OL_w$ starts off directly assuming that a representation of the form above exists, or rather an equivalent one in terms of the $\fl_w$ and $\gl_w$ kernels\footnote{Actually, the two representations are not quite equivalent: the representation of the functional action with $f,g$ is more generally valid than that with $h$, since the former might be finite where the latter diverges.} and formulates a set of equations with boundary conditions that these kernels must satisfy. In particular $\gl_w$ is fixed in terms of $\fl_w$ and the latter satisfies the equation
\bea
\mathcal R_z \fl_w(z)-(1-z)^{2\Df-2}\fl_w(\mbox{$\frac 1{1-z}$})-z^{2\Df-2}\fl_w(\mbox{$\frac 1{z}$})=\delta(z-w)+\delta(1-z-w)\label{eq:eqnf2}
\eea
subject to boundary conditions spelled out in \reef{eq:boundcond}. We have shown that a solution to these equations satisfying the right boundary conditions can be constructed numerically for general $\Df$ (and exactly for special values, see \ref{app:funcsol}). We can now ask if this definition of the functional is compatible with the first one. Concretely we can discuss whether the solutions constructed in this way admit a representation of the form:
\bea
\fl_w(z)&\overset{?}{=}-\sum_{n=0}^{\infty}\left[ G_{\Delta_n^F}(w|\Df) f_{\alpha_n^F}(z)+\partial_{\Delta}G_{\Delta_n^F}(w|\Df) f_{\beta_n^F}(z)\right]\,.\label{eq:expf2}
\eea
This can indeed be checked explicitly for all cases where we can find $\fl_w(z)$ exactly. More generally, we would need to show that for sufficiently small $w$ of the form $\fl_w(z)\sim a_n(z) w^n+b_n(z) \log(w) w^n$. We cannot establish this rigorously at this point, since in general we can only solve for the kernels numerically using the integral equation of the previous subsection. All we can say is that if this is true then the coefficients in such an expansion must satisfy the equation \reef{eq:eqnf2} without the $\delta$ function terms, and the solutions of that homogeneous equation are precisely the basis functionals which have been constructed for all $\Df$. Furthermore we can argue that the $\delta$ function terms can arise from the full sum. Let us take the limit $w,-z\to 0^+$ in the expression above. We find
\bea
\fl_w(z)&\underset{w,-z\to 0^+} \sim -\sum_{n=0}^\infty \frac{2 w}{\pi^2} \partial_n \left(\frac{w}{z}\right)^{2n}=\frac{2w}{\pi^2} \frac{\log(-w/z)}{z^2-w^2}
\eea
It is easy to check that this satisfies \reef{eq:eqnf2} in the same limit.

An analogous argument tells us that the functional action can be similarly expanded with coefficients which can be identified with functionals $\alpha_n(\Delta)$ and $\beta_n(\Delta)$, therefore recovering our first definition of the functional. It would be important to establish this rigorously.

\section{Dispersion formula}
\label{sec:dispersion}
\subsection{The Polyakov bootstrap and completeness}
\label{sec:polycomplete}

In the preceding sections we have constructed master functionals $\OLU_w$ via their kernels $\ful_w$ and $\gul_w$, and have also argued that it should be possible to express these functionals in terms of the 1d functional bases. Here we will be interested in the functional {\em actions} $\OLU_w(
\Delta)$. Let us begin by recalling that the functionals satisfy the property:
\bea
\OL_{1-w}(\Delta)-\OL_{w}(\Delta)=F_{\Delta}(w)=\OU_{w}(\Delta)-\OU_{1-w}(\Delta)\,,\label{eq:idid}
\eea 
This motivates the following definitions:
\bea
\mathcal P^F_\Delta(w|\Df)&:=\OL_w(\Delta|\Df)+G_{\Delta}(w|\Df)\,,\\
-\mathcal P^B_\Delta(w|\Df)&:=\OU_w(\Delta|\Df)-G_{\Delta}(w|\Df)\,.\label{eq:funcpolyakov}
\eea
We call $\mathcal P^B_\Delta$ and $
\mathcal P^F_\Delta$ respectively the bosonic and fermionic {\em Polyakov blocks}. Note that by construction they are crossing symmetric functions. Using the expansions of the master functionals in the 1d functional bases leads to a more explicit representation of these objects:
\bea
\mathcal P_\Delta^{B,F}(w|\Df)&=\mathcal P^{B,F}_\Delta(1-w|\Df)\\
&=G_{\Delta}(w|\Df)-\sum_{n=0}^{+\infty} \left[G_{\Delta_n^{B,F}}(w|\Df) \alpha_n^{B,F}(\Delta)+\partial_\Delta G_{\Delta_n^{B,F}}(w|\Df) \beta_n^{B,F}(\Delta)\right]\label{eq:polyakov}
\eea
The above amount to the conformal block decompositions of the Polyakov blocks, and we see they are determined by the  functional actions $\alpha_n(\Delta), \beta_n(\Delta)$. These expressions are consistent with the interpretation of the Polyakov blocks as particular crossing symmetric sums of Witten exchange diagrams in AdS$_2$ \cite{Mazac2019a}. Thus from this perspective the Polyakov blocks provide us with an independent way of computing functional actions $\OLU_w(\Delta)$. In this work we will do the reverse and compute the blocks using the master functionals.

The master functionals are crossing compatible, and hence they can be applied to the crossing equation to derive sum rules on the OPE data. This has important consequences. Let us restrict the discussion to $
\OL_w$, with analogous results holding for $\OU_w$. The first consequence is that the constraints of crossing symmetry are fully equivalent to the vanishing of the $\OL_w$ sum rules:
\begin{eqnarray}
\boxed{
\sum_{\Delta} a_{\Delta} F_\Delta(z)=0 \qquad \forall z \in(0,1) \quad \Leftrightarrow \qquad  \sum_{\Delta} a_{\Delta} \OL_w(\Delta)=0\qquad \forall w \in(0,1)
}
\end{eqnarray}
The proof of this statement is straightforward. To prove the implication $\Rightarrow$ we simply apply $\OL_w$ to the crossing equation and use crossing compatibility. As for the implication $\Leftarrow$ it follows from antisymmetrizing the $
\OL_w$ sum rule in $w$ and using
\reef{eq:idid}. The conclusion is that the set of $\OL_w$ for $w\in(0,1)$ forms a new, complete, basis of functionals for the crossing equation.

If we use the definition of the Polyakov block the corresponding sum rules can be written as:
\begin{eqnarray}
\boxed{
\sum_{\Delta} a_{\Delta} \OL_w(\Delta)=0\qquad \Leftrightarrow \qquad \sum_{\Delta} a_{\Delta} \mathcal P^F_\Delta(w)=\sum_{\Delta} a_{\Delta} G_{\Delta}(w|\Df)} \label{eq:polyakovbs}
\end{eqnarray}
The righthand equation is what is known as the Polyakov bootstrap: it states that for solutions to crossing we may replace conformal blocks in the OPE expansion of correlation functions by crossing-symmetric Polyakov blocks. Using our previous result, we can now conclude that the constraints of crossing symmetry are equivalent to those of the Polyakov bootstrap. Note that for this proof we have relied only on the existence of the functional $\OL_w$. We did not need an independent definition of the Polyakov block other than through the action $\OL_w(\Delta)$. 

\paragraph{Completeness} Let us make a few (incomplete) remarks on what we mean by completeness' of functional bases. Usually what we mean is simply the constraints of crossing symmetry are fully captured by acting with all functionals in the basis. Clearly every functional provides a necessary condition for crossing. It is establishing sufficiency which is difficult. In \cite{Mazac2019a} this was shown for both bosonic and fermionic functional bases by proving:
\bea
\left(\sum_{\Delta}\sum_{n=0}^{\infty}-\sum_{n=0}^{\infty}\sum_{\Delta}\right)\left[G_{\Delta_n^F}(w) a_{\Delta} \alpha_n^F(\Delta)+\partial_\Delta G_{\Delta_n^F}(w) a_{\Delta} \beta_n^F(\Delta)\right]=0\,.\label{eq:commute}
\eea
Indeed it is easy to see by using the OPE that the Polyakov bootstrap conditions are satisfied if the equation above holds together with all sum rules of $\alpha_n,\beta_n$.

This was proven by using the detailed form of the functional actions, together with upper bounds on the OPE coefficients. In the present work however, we can recognize the above as simply the condition of crossing compatibility for the $\OL_w$ functional, which we can simply verify explicitly given the functional kernels. So it may seem that we have managed to bypass a lot of work with this new approach, but actually this is not correct. For instance, if we are given the explicit $\fl_w$, $\gl_w$ kernels satisfying crossing compatibility, it is not obvious how to prove the validity of the representation of $\OL_w$ in the functional basis, as we've already discussed in subsection \ref{sec:subtle}. Alternatively, if we start off defining the $
\OL_w$ by that representation, then it is proving crossing-compatibility that is non-trivial. 

In the above, completeness referred to the full encoding of the constraints of crossing symmetry. But we can perhaps aim for more general results. Using the expansion of $\OL_w$ in the 1d functional basis we have
\bea
\OL_w-\OL_{1-w}=\mathcal E_w \Rightarrow \mathcal F(w)=\sum_{n=0}^{+\infty}\left(F_{\Delta_n^F}(w) \alpha_n^F[\mathcal F]+\partial_\Delta F_{\Delta_n^F}(w) \beta_n^F[\mathcal F]\right) \label{eq:completenessF}
\eea
This reads like a decomposition of the test function $\mathcal F$ in a ``basis'' formed by the~$F_{\Delta_n^F}, \partial_\Delta F_{\Delta_n^F}$. In general it is a difficult question to determine for which kinds of test functions the above holds, but we can try to make an educated guess. We choose $\mathcal F(z)=-\mathcal F(1-z)$ which are analyic in $\mathbb C\backslash(-\infty,0]\cup[1,\infty)$ and satisfy 
\bea
\mathcal F(c+it)\underset{t\to \infty}=O(t^{1-\epsilon})\, \qquad \mathcal F(1-t)\underset{|t|\to 0^+}=O(|t|^{-2\Df-1+\epsilon})\,, \qquad \epsilon>0\,.\label{eq:boundarycondsF}
\eea
The conditions were chosen such that $\OL_w[\mathcal F], \alpha_n^F[\mathcal F], \beta_n^F[\mathcal F]$ are all finite. Note in particular that these conditions hold for arbitrary finite linear combinations of crossing vectors $F_{\Delta}$ with $
\Delta\geq 0$, for which we know that equation \reef{eq:completenessF} holds\footnote{In that case it encodes crossing symmetry of the Polyakov block, which we know is true by direct computation of the latter as a crossing-symmetric sum of Witten exchange diagrams.}. The completeness statement \reef{eq:completenessF} is much stronger than the statement \reef{eq:completeness} of the equivalence of the crossing equation with the functional basis sum rules. Indeed that equivalence follows trivially from \reef{eq:completenessF} by choosing %
\bea
\mathcal F(w)=\sum_{\Delta} a_{\Delta} F_{\Delta}(w)
\eea
with the same boundary conditions \reef{eq:boundarycondsF}. 

It would be very interesting to develop these arguments more carefully to understand the mathematically rigorous sense in which the crossing vectors form a basis for some space.

\subsection{From functionals to dispersion relations}

We have seen that master functional actions are intimately related to Polyakov blocks via equations \reef{eq:funcpolyakov}. On the other hand, we have also seen a different representation for the functional action expressed through \reef{eq:funcactions}. Combining these two observations, we can arrive at new expressions for the Polyakov blocks. To make these expressions more suggestive, let us first introduce the bosonic and fermionic double discontinuities, respectively:
\bea
\mbox{dDisc}_B\, \mathcal G(z)&:= \mathcal G(z)-\mathcal R_z \left[\mathcal G(\mbox{$\frac z{z-1}$})\right] (1-z)^{-2\Df}\,,&\qquad z&\in (0,1)\\
\mbox{dDisc}_F\, \mathcal G(z)&:= \mathcal G(z)+\mathcal R_z \left[\mathcal G(\mbox{$\frac z{z-1}$})\right] (1-z)^{-2\Df}\,,&\qquad z&\in (0,1)\,.
\eea
When no subscript is indicated, we will always mean the bosonic case. These definitions imply:
\bea
\mbox{dDisc}_B\, \mathcal P^B_{\Delta}(z|\Df)=\mbox{dDisc}_B\, G_{\Delta}(z|\Df)=2\sin^2\left[\frac \pi 2 (\Delta-2\Df)\right]G_{\Delta}(z|\Df)\,,\\
\mbox{dDisc}_F\, \mathcal P^F_{\Delta}(z|\Df)=\mbox{dDisc}_F\, G_{\Delta}(z|\Df)=2\cos^2\left[\frac \pi 2 (\Delta-2\Df)\right]G_{\Delta}(z|\Df)\,.
\eea
We can now use \reef{eq:funcactions} to find:
\bea
\label{eq:polyposrep}
\mathcal P_\Delta^F(w|\Df)&=+\int_0^1\ud z\, \ghl_w(z|\Df) \mbox{dDisc}_F\, G_{\Delta}(z|\Df)& \qquad \Delta&>2\Df-1\,,\\
\mathcal P_\Delta^B(w|\Df)&=-\int_0^1\ud z\, \ghu_w(z|\Df) \mbox{dDisc}_B\, G_{\Delta}(z|\Df)& \qquad \Delta&>2\Df\,.
\eea
We can now go a step further and use these expressions to rewrite the Polyakov bootstrap statement in \reef{eq:polyakovbs}. To do this we first define subtracted correlators:
\bea
\underline{\mathcal G}(z)&:=\mathcal G(z)-\sum_{0\leq\Delta\leq 2\Df-1}\!\!\!\! a_{\Delta} \mathcal P^F_{\Delta}(z)\,,\\
\overline{\mathcal G}(z)&:=\mathcal G(z)-\sum_{0\leq\Delta\leq 2\Df} a_{\Delta} \mathcal P^B_{\Delta}(z)\,.
\eea
Note that in these definitions there is always at least one subtraction corresponding to the identity Polyakov blocks. In passing we note that these are the same as the generalized free field correlators,\footnote{This follows from $\beta_n^{B,F}(0)=0\,, \alpha_n^{B,F}(0)=-a_{\Delta_n^{B,F}}^{\mbox{\tiny gff}}$.}
\bea
\mathcal P_0^{B,F}(w|\Df)=\mathcal G^{B,F}(w)\,.
\eea
The statements $\sum_{\Delta} a_{\Delta} \mathcal P_{\Delta}=\sum_{\Delta} a_{\Delta} G_{\Delta}=\mathcal G$ become the following crossing-symmetric CFT dispersion relations:
\begin{subequations}
\label{eq:dispersions}
\begin{empheq}[box=\fbox]{align}
\underline{\mathcal G}(w)&=+\int_0^1\ud z \, \ghl_w(z|\Df)\, \mbox{dDisc}_F\left[\underline{\mathcal G}(z)\right]\\
\overline{\mathcal G}(w)&=-\int_0^1\ud z \, \ghu_w(z|\Df)\, \mbox{dDisc}_B\left[\overline{\mathcal G}(z)\right]
\end{empheq}
\end{subequations}
This is the main result of this section, and perhaps of this entire paper. It tells us that general CFT correlators with $z=\bar z$ can be computed from their double discontinuity, up to a finite set of contributions of low dimension operators. Remarkably, the validity of the dispersion relations is equivalent to the sum rules for the master functionals, e.g.
\bea
\sum_{\Delta} a_{\Delta} \OL_w(\Delta)=0 \qquad \Leftrightarrow\qquad  \underline{\mathcal G}(w)&=\int_0^1\ud z \, \ghl_w(z|\Df)\, \mbox{dDisc}_F\left[\underline{\mathcal G}(z)\right]
\eea
and hence satisfying the dispersion relation for $w\in(0,1)$ is equivalent to solving the crossing equation. 

The dispersions relations are manifestly crossing symmetric, since the master funcional kernels definitely are. As such the content of the crossing symmetry constraints here really is the statement that once we put some function $\mathcal G(z)$ under the integral sign we should get back the same thing. For instance, this is not true for any single conformal block, since after integration we get out a (crossing-symmetric) Polyakov block of the same dimension. However, it is true for arbitrary linear combinations of Polyakov blocks, including finite ones, as those are valid (though non-unitary) solutions to crossing.

\subsection{From dispersion relations to functionals}
\label{sec:dispderivation}
We have seen how the master functionals $\OLU_w$ lead to dispersion relations for CFT correlators. Our goal now is to reverse the logic and show that by demanding that a  dispersion relation exists is so constraining that it uniquely fixes the functional kernels. Furthermore, we will show that the dispersion relations can be used to extract the sum rules associated to the bases functionals, thus closing the circle. For concreteness we will mostly focus on the bosonic case. 

Let us start therefore by assuming a dispersion relation exists of the form
\bea
\mathcal G(w)=\int_0^1 \ud z\, g(w,z) \mbox{dDisc}[\mathcal G(z)]
\eea
with $\mbox{dDisc}=\mbox{dDisc}_B$, and for clarity we have set $
\ghu_w(z)\equiv g(w,z)$. 

For crossing to hold we must have that $g(w,z)=g(1-w,z)$. The dispersion relation implies
\bea
\mbox{dDisc}_w g(w,z)\equiv g(w,z)-(1-w)^{-2\Df} \mathcal R_w\left[ g\left(\mbox{$\frac w{w-1}$},z\right)\right]=\delta(w-z) \label{eq:ddiscK}
\eea
for $w,z \in (0,1)$. The idea now is that for a certain choice of boundary conditions this equation has a unique solution. The strategy is similar to the derivation of the integral equation satisfied by $f_w(z)$ in section \ref{sec:fredholm}, but now we work with the $w$ variable instead of $z$ (and with the $g_w(z)$ kernels instead of $f_w(z)$, although this could be changed trivially). We introduce
\bea
\hat g(w,z)=\frac{g(w,z)}{\sqrt{w(1-w)}}
\eea
Changing variables \reef{eq:ddiscK} becomes:
\bea
\mathcal I_w \hat g(w,z)=w^{-2\Df-\frac 32} \hat g\left(\mbox{$\frac {w-1}{w}$},z\right)-\frac{(1-z)^{2\Df-\frac 12}}{\sqrt{z(1-z)}}\,\delta\left(w-\mbox{$\frac{1}{1-z}$}\right)\,,\qquad w>1\,.
\eea
We now use the Cauchy formula for $\hat g$ and deform the contour to pick up the discontinuities along $w<0$ and $w>1$. In terms of $g_w(z)$ this gives
$$
\label{eq:dispg}
\boxed{
g_w(z)=\tilde K_{\Df}(w,z)- \int_0^1 \ud w' \tilde K_{\Df}(w,w') g_{w'}(z) 
}
$$
with
\bea
\tilde K_{\Df}(w,w')=\frac{1}{\pi}\sqrt{\frac{w(1-w)}{w'(1-w')}} \frac{1+w'}{\left(w-\frac{w'}{w'-1}\right)\left(\frac 1{1-w'}-w\right)}\,(1-w')^{2\Df-\frac 32}\,.
\eea
In terms of $\fu_w(z)=z^{2\Df-2}g_w(1-1/z)$ the above becomes
\bea
\fu_w(z)=\frac{2}{\pi}\sqrt{\frac{w(1-w)}{z(z-1)}} \frac{z-\frac 12}{(w-z)(1-w-z)}-\int_0^{1} \ud w' \tilde K_\Df(w,w') f_{w'}(z)\,,
\eea
This is reminiscent of the integral equation \reef{eq:dispf}, but now the integration runs over $w$ instead of $z$. Notice we have the relation
\bea
\tilde K_{\Df}\left(\mbox{$\frac{z'-1}{z'},\frac{z-1}z$}\right)=(z')^{2\Df} z^{2-2\Df} K_{\Df}(z,z')\,.
\eea
The present integral equation can be solved numerically just as easily as \reef{eq:dispf}, and it can be checked that both equations lead to the same result for the kernels. Hence we have succeeded in deriving the functional kernels from the existence of a dispersion relation. 

\paragraph{Deriving functional sum rules}

To close the circle, let us now show that the dispersion relation explicitly leads directly to the infinite set of sum rules associated to the $\alpha_n, \beta_n$ functionals. For defineteness we work with the dispersion relation associated to $\OL_w$ which we recall here:
\bea
\underline{\mathcal G}(w)&=\int_0^1\ud z \, \gl_w(z|\Df)\, \mbox{dDisc}_F\left[\underline{\mathcal G}(z)\right]\,, \quad 
\underline{\mathcal G}(w)\equiv \mathcal G(w)-\sum_{0\leq\Delta\leq 2\Df-1}\!\!\!\! a_{\Delta} \mathcal P^F_{\Delta}(w)
\eea
Firstly, note that
\begin{multline}
\mbox{dDisc}_{F;w} \gl_w(z)=\delta(w-z)\\ \Rightarrow \gl_w(z)=\sum_{n=0}^{\infty}\left[g_{\alpha_n^F}(z) G_{\Delta_n^F}(w|\Df)+g_{\beta_n^F}(z) \partial_{\Delta} G_{\Delta_n^F}(w|\Df)\right]
\end{multline}
where, for now, $g_{\alpha_n^F}(z)$ and $g_{\beta_n^F}(z)$ are defined by the above expansion. The expansion hold away from $w=z$, by expanding in a basis of functions with zero double discontinuity, namely the blocks and their derivatives with $\Delta=\Delta_n^F$.

The idea now is to match powers of $w$ on both sides of the dispersion relation. The integral on the righthand side leads in general to two sets of terms. Away from the small $z$ integration region the above expansion of $\gl_w(z)$ is valid and leads to a set of ``analytic'' terms of the form $w^n$, $w^n \log(w)$. At the same time the small $z\sim w$ integration region reproduces ``non-analytic'' terms in $w$, i.e. with non-zero double discontinuity. These must match those in the expansion of $\mathcal G(w)$ on the lefthand side. 

Sum rules arise by demanding that the $w^n \log(w)$ and $w^n$ terms cancel out. To make sure that these terms dominate the small $w$ expansion we must do enough subtractions on the correlator to guarantee that those terms are leading. Define therefore
\bea
\underline{\mathcal G}_n(w) := \mathcal G(w)-\sum_{0\leq\Delta\leq \Delta_n^F}\!\!\!\! a_{\Delta} \mathcal P^F_{\Delta}(w)
\eea
Let us match terms on both sides of the dispersion relation applied to $\underline{\mathcal G}_n(w)$. Using the conformal block expansion of $\mathcal P^F_{\Delta}$ we have
\begin{multline}
\underline{\mathcal G}_n(w)\underset{w\to 0}=\sum_{m=0}^n \left[ G_{\Delta_n^F}(w)\left(\sum_{0\leq \Delta\leq \Delta_m^F} a_{\Delta}\alpha_n^F(\Delta)\right)\right.\\
\left.+\partial_{\Delta} G_{\Delta_n^F}(w)\left(\sum_{0\leq \Delta\leq \Delta_n^F} a_{\Delta}\beta_m^F(\Delta)\right)\right] + O(w^n \log(w))
\end{multline}
On the other hand expanding $\gl_w$ will lead to terms such as for instance
\begin{multline}
\int_0^1 g_{\alpha_m^F}(z) \mbox{dDisc}_F \left[\underline{\mathcal G}_n\right]=\int_0^1 g_{\alpha_m^F}(z) \mbox{dDisc}_F \left[\sum_{\Delta\geq \Delta_n^F} a_{\Delta} G_{\Delta}(z)\right]
:=\sum_{\Delta\geq \Delta_n^F} a_\Delta \alpha_m^F(\Delta)
\end{multline}
Therefore matching powers of $w$ leads indeed to the sum rules
\bea
\sum_{\Delta\geq 0} a_{\Delta}\alpha_n^F(\Delta)=0\, \qquad \sum_{\Delta\geq 0} a_{\Delta}\beta_n^F(\Delta)=0\,.
\eea
There is seemingly a gap missing in this derivation related to the definition of the Polyakov block for $\Delta\leq \Delta_n^F$ but this is easy to fix. We simply define this object as the action of the dispersion equation on a single conformal block.\footnote{This defines $\mathcal P_\Delta^F$ for $\Delta\geq 2\Df-1$ but smaller $\Delta$ can be obtained easily via analytic continuation, by e.g. subtractions.} From this definition its conformal block decomposition can be computed from that equation, proving incidentally that the coefficients $g_{\alpha_n^F}(z)$, $g_{\beta_n^F}(z)$ are the functional kernels which compute the actions of $\alpha_n^F$ and $\beta_n^F$.

\section{Application: Correlator Bounds}
\label{sec:appbounds}
It is only appropriate that our first application of the dispersion relations is to prove bounds on CFT correlators, since after all, the search for such bounds was what originally motivated our definitions of the master functionals in section \ref{sec:extfuncs}. The bounds follow almost immediately from the assumption that the kernels $\ghul_w(z)$ are positive for $z,w\in(0,1)$ the evidence for which was shown in figure \ref{fig:kernels}. Since the double discontinuities of the subtracted correlators are manifestly positive this proves that:
\bea
\underline{\mathcal G}(w)\geq 0\,, \qquad \overline{\mathcal G}(w)\leq 0\,, \qquad \mbox{for}\quad w\in (0,1)
\eea
or equivalently:
\begin{subequations}
\label{eq:corrbounds}
\begin{empheq}[box=\fbox]{align}
\mathcal G(w)&\geq \mathcal G^F(w)+\sum_{0<\Delta\leq 2\Df-1}\!\!\!\! a_{\Delta} \mathcal P^F_{\Delta}(w)\,,\\
\mathcal G(w)&\leq \mathcal G^B(w)+\sum_{0<\Delta\leq 2\Df} a_{\Delta} \mathcal P^B_{\Delta}(w)\,.
\end{empheq}
\end{subequations}

It is important to emphasize that these bounds hold for any CFT correlator in any spacetime dimension restricted to the line $z=\bar z$. In particular, the upper bound does {\em not} require any particular assumptions on the spectrum other than unitarity. If we do assume a gap in the spectrum $\Delta_g\geq 2\Df$ it correctly reduces to our original proposed bound $\mathcal G(w)\leq \mathcal G^B(w)$. If we do not make this assumption, the bound is still valid with the caveat that we do not claim that it is optimal. 

These bounds are in fact {\em stronger} than our initial aim of proving
\bea
\mathcal G^F(w)\leq\mathcal G(w)\leq \mathcal G^B(w)\,,
\eea
To appreciate this it is important to point out that we find
\bea
\mathcal P_{\Delta}^F(w|\Df)&\geq 0\,,& \qquad 0&\leq \Delta\,,\\
\mathcal P_{\Delta}^B(w|\Df)&\geq 0\,,& \qquad 0&\leq \Delta\leq 2\Df\,,\label{eq:pospoly}
\eea
for all $\Df$. While the first statement is equivalent to the positivity requirements \reef{eq:poslower} on $\OL_w(\Delta)$, required to obtain a valid, gap independent, lower bound, the second statement is a new observation. Recall that, as follows from the representations \reef{eq:polyposrep},
the issue is that positivity of the kernels $
\gul_w$ is not sufficient to determine that of the Polyakov blocks for sufficiently small $\Delta$. However, the positivity properties above can be checked explicitly on a case by case basis by numerically evaluating the master functional  actions $\OLU_w(\Delta)$. In this way we have checked numerically for many $\Df$ that the positivity properties above are indeed true. 

The positivity properties can also be understood analytically from the small $w$ expansions of the master functionals. Indeed, for small enough $w$ we have
\bea
\OL_w&\underset{w\to 0}{\sim}- w  \log(w) \left( \beta_0^F+\ldots\right)\,,\\
\OU_w&\underset{w\to 0}{\sim}     \alpha_0^B+\ldots\,,
\eea
In passing, note that these expressions are nicely consistent with the expected link between the correlator minimization/maximizion problems and the gap maximization/ope maximization problems respectively as outlined in section \ref{sec:extfuncs}. Indeed, the functionals $\beta_0^F$ and $\alpha_0^B$ are precisely the extremal functionals corresponding to these problems in $d=1$ \cite{Mazac:2016qev,Mazac:2018}. In what respects the Polyakov blocks we have:
\footnote{A subtlety is that for $\OU_w$ the limits $\Delta\to \Delta_n^B$ and $w\to 0$ don't commute, since $\alpha_0^B(\Delta_n^B)\propto \delta_{n,0}$. Close to $\Delta=\Delta_n$ other terms in the expansion of $\OU_w$ dominate in the small $w$ limit, and this is important to have positivity in the final result.}
\bea
\label{eq:positivitysmallw}
\mathcal P^F_\Delta(w)=\OL_w(\Delta)+G_\Delta(w|\Df) &\underset{w\to 0}{\sim}\left\{ 
\begin{array}{cc}
-w \log(w) \beta_0^F(\Delta)\geq 0\,,& \Delta\geq 2\Df+1\\
w^{\Delta-2\Df} \geq 0\,, & \Delta< 2\Df+1
\end{array}
\right.
\\
-\mathcal P^B_\Delta(w)=\OU_w(\Delta)-G_\Delta(w|\Df) &\underset{w\to 0}{\sim}\left\{ 
\begin{array}{cc}
\alpha_0^B(\Delta)\geq 0\,,& \Delta>2\Df\\
-w^{\Delta-2\Df} \leq 0\,, & \Delta< 2\Df
\end{array}
\right.
\eea
in perfect agreement with \reef{eq:pospoly}. 

Let us conclude this section with some small observations. We can reformulate our bounds in terms of the {\em non-gaussianity} $Q_z:=\mathcal G(z)/\mathcal G^B(z)$ of a CFT correlator \cite{Rychkov:2016mrc}. The upper bound becomes
\bea
Q_z \leq 1\,, \qquad \Delta_g\geq 2\Df
\eea
Our result therefore generalizes what is known as the Leibowitz inequality for the 3d Ising model any CFT correlator with the right gap, albeit restricted to a line. As for the lower bound, which is independent of any assumptions on the spectrum, is:
\bea
\frac 13\leq \frac{1-\left(\frac 12\right)^{1+2\Df}}{1+\left(\frac 12\right)^{1+2\Df}}\leq \frac{z^{-2\Df}+(1-z)^{-2\Df}-1}{z^{-2\Df}+(1-z)^{-2\Df}+1}\leq Q_z
\eea
As $\Df$ increases we see the lower bound on $Q_z$ rapidly approaches one. This bound is not easily generalizable away from $z=\bar z$. However, identifying $z$ in the above with $\sqrt{z\bar z}$ for general Euclidean kinematics we expect it should approximately hold, at least for $z\bar z\ll 1$, as should the upper bound. 

\section{Application: Regge physics}
\label{sec:regge}
In this section we will discuss we will study the implications of the dispersion relations for the Regge limit of a CFT correlator.

\subsection{Definitions and properties}

The Regge limit in a certain OPE channel of a CFT correlator is defined as that channel's OPE limit, but taken after an analytic continuation to the second sheet \cite{Costa:2012cb}\footnote{See also the works \cite{Kravchuk:2018htv,Caron-Huot:2020nem} for formal aspects of the Regge limit.}
. Here we will focus on correlators of identical bosonic scalar operators and consider the $t$-channel Regge limit in the forward limit, i.e. setting $\bar z=z$ after going to the second sheet. We define:
\bea
\, (1-z)^{2\Df} \frac{\left[ \mathcal G(e^{2\pi i} z, z)-\mathcal G(z,z)\right]}{i} \underset{z\to 1}\equiv 
\mu \, (1-z)^{1-j_0}\,.
\eea
with $\mu \in \mathbb C$ and $j_0$ a real number.
The above is analogous (in fact, somewhat more than analogous) to a high energy limit of a scattering amplitude. In spite of several important results, there remain many questions about this limit for generic non-perturbative CFTs. For instance, one does not even know if the power-law behaviour is the only one allowed. 
 In this section we will simply assume the behaviour above but in fact our discussion can be easily modified without much effort to more general asymptotic behaviours.  

There are two simple facts we know about $\mu$ and $j_0$. Consider the $s$-channel OPE decomposition of the expression between brackets above:
\bea
\mathcal G(e^{2\pi i} z, z)-\mathcal G(z,z)=\sum_{\Delta,\ell} a_{\Delta,\ell} \left(e^{i\pi(\Delta-2\Df)}-1\right) G_{\Delta,\ell}(z,z|\Df)
\eea
It follows that $|\mathcal G(e^{2\pi i} z, z)|\leq \mathcal G(z,z)$ for real $z<1$ and hence we must have $j_0\leq 1$ and, in the case $j_0=1$, $|\mu|\leq 2$.

The Regge behaviour of the correlator is 
closely related to that of its double discontinuity. First let us define:
\bea
(1-z)^{2\Df} \mbox{dDisc}_s \mathcal G(z,z)
\underset{z\to 1}{\sim} \nu (1-z)^{1-\tj}
\eea
From the OPE decomposition we have
\bea
(1-z)^{2\Df} \mbox{dDisc}_s \mathcal G(z,z)=
(1-z)^{2\Df}\sum_{\Delta,\ell} a_{\Delta,\ell}\,2 \sin^2\left[\frac{\pi}2(\Delta-2\Df)\right] G_{\Delta,\ell}(z,z|\Df)]\,.
\eea
From this it follows that: %
\bea
\mbox{Im}\, \mu\neq 0:& \qquad j_0= \tj\quad \mbox{and}\quad \nu=\mbox{Im}\mu\geq 0\\
\mbox{Im}\, \mu=0:&\qquad j_0> \tj\quad \mbox{and}\quad \nu\geq 0 \label{eq:relj0}
\eea
In fact, as we will see below, we will be able to more precisely characterize those situations where $j_0$ is or is not equal to $\tj$. For now, let us merely note that bounding $\tj$ from below also bounds $j_0$. 

\paragraph{Regge theory} Before we study what more can be said about $j_0$, let us briefly make contact with the more usual approach of studying the Regge limit, and in particular the study of leading Regge trajectories.  The starting point is the basic assumption that the contribution of a single Regge pole, capturing the contribution of the leading Regge trajectory, dominates in the Regge limit. It can then be shown that \cite{Costa:2012cb,Kravchuk:2018htv}:
\bea
(1-z)^{2\Df}\,\frac{\left[ \mathcal G(e^{2\pi i} z, z)-\mathcal G(z,z)\right]}i \underset{z \to 1}= \int_{-\infty}^{+\infty} \frac{\ud c}{2\pi i} (1-z)^{1-J_0(\Delta)} C_{\Delta,J_0(\Delta)}  \Bigg |_{\Delta=\frac d2+i c}\label{eq:reggecorr}
\eea
where $J_0(\Delta)$ is the leading Regge trajectory and $C_{\Delta,J_0(\Delta)}$ is related to the Euclidean OPE density. This gives us:
\bea
j_0\leq J_0 \qquad (:= J_0(d/2) )\,.
\eea
The parameter $J_0$ is called the intercept. This inequality implies that if $J_0<1$ then the same is true of $j_0$. Conversely, $j_0=1$ implies $J_0=1$. This inequality does not follow immediately from \reef{eq:reggecorr} and requires some explanation.\footnote{We thank S. Caron-Huot for communication regarding these points, including the bounds on $J_0$ arising from convexity.} The leading Regge pole trajectory is determined by the breakdown in convergence of the Lorentzian inversion formula \cite{Caron-Huot:2017vep}. This implies not only upward concavity of $J_0(\Delta)$ for real $\Delta$ but also that $\mbox{Re} J_0(\Delta)\leq J_0(\mbox{Re} \,\Delta)$ \cite{Caron-Huot:2020ouj} (see also \cite{Komargodski2013}) and hence $J_0\geq \mbox{Re}\, J_0\left(\frac d2+i c\right)$. This leads to the above result. In passing we mention that upward concavity of $J_0(\Delta)$ leads to lower bounds on $J_0$ under moderate assumptions. Assuming the leading Regge trajectory asymptotes to a set of operators of twist $\tau\sim 2\Df$ gives
\bea
J_0 \geq \frac d2-2\Df\,.
\eea
If we furthermore specify that the trajectory passes through an operator of twist $\tau_2$ and spin $2$ we get the stronger\footnote{More general bounds are possible. For any two operators in the leading Regge trajectory concavity implies
\bea
J_0(d/2)\geq \ell_1+\left(\frac d2-\Delta_1\right) \frac{\ell_2-\ell_1}{\Delta_2-\Delta_1}\,.
\eea
}
\bea
J_0 \geq \frac{d}2-\tau_2 \qquad \left(= 2-\frac d2\quad \mbox{for}\quad \tau_2=d-2\right).
\eea
These bounds are reminiscent of the ones we will find below. We emphasize however that our bounds on $j_0$ will not really assume anything particular about the spectrum.

Since $j_0$ captures the damping effect of the complex $\Delta$ integral, in general we expect that it should be quite different from $J_0$. Indeed, to better understand the difference between $j_0$ and $J_0$ note that the Lorentzian inversion formula tells us $J_0$ is really a probe of the lightcone limit of the $s$-channel double discontinuity, which receives contributions from operators of fixed twist but large spin. In contrast, $j_0$ probes the $t$-channel limit of the double discontinuity, which is sensitive to  operators of large dimension and arbitrary spin. 

\paragraph{Connection to high energy properties} Let us try to make this last statement more quantitative. Let us begin by noting that it is always possible to choose $z_0<1$ such that
\bea 
\int_{z_0}^1 \ud z (1-z)^{2\Df-1-\alpha} \mbox{dDisc}\, \mathcal G(z)<\infty \Leftrightarrow 1-\tj>\alpha
\eea
Notice that the same statement can be made even after  subtracting an arbitrarily large number of conformal blocks from $\mathcal G(z)$. When the integral converges we can therefore use the OPE to find
\begin{multline}
\int_{z_0}^1 \ud z (1-z)^{2\Df-1-\alpha} \mbox{dDisc} \,\left[\mathcal G(z)-\sum_{\Delta\leq \Delta^*} a_{\Delta} G_{\Delta}(z|\Df)\right]\\=2\,\sum_{\Delta> \Delta^*} a_{\Delta} \sin^2\left[\frac{\pi(\Delta-2\Df)}2\right]\int_{z_0}^1 \ud z\, (1-z)^{2\Df-1-\alpha} G_{\Delta}(z|\Df)<\infty
\end{multline}
Let us choose $\Delta^*\gg 1$. In the limit $z \to 1$ keeping $\Delta \sqrt{1-z}$ fixed the conformal block simplifies and the above turns into\footnote{The precise statement is that in this limit we have $G_{\Delta}(z) \sim  \frac{2\Gamma(2\Delta)}{\Gamma(\Delta)^2 }K_0(2\Delta \sqrt{1-z})$.}
\bea
\sum_{\Delta>\Delta^*} \left(\frac{a_{\Delta}}{a_\Delta^{\mbox{\tiny free}}}\right) \frac{2 \sin^2\left[\frac{\pi(\Delta-2\Df)}2\right]}{\Delta^{1-2\alpha}} <\infty \Leftrightarrow 1-\tj>\alpha
\eea
This condition can be rewritten more elegantly by introducing the average squared anomalous dimension:
\begin{eqnarray}
\boxed{
\overline{\gamma_n^2}\underset{n\to \infty}{\simeq}  \Delta_n^{-2(1-\tj)}\,, \qquad \overline{\gamma_n^2}:=\frac{1}{C_n} \sum_{|\Delta-\Delta_n|\leq C_n} \left(\frac{a_{\Delta}}{a_\Delta^{\mbox{\tiny free}}}\right) \frac{4 \sin^2\left[\frac{\pi(\Delta-2\Df)}2\right]}{\pi^2}
}
\end{eqnarray}
where $\Delta_n=2\Df+2n$ as usual. The above establishes a relation between the Regge limit and the properties of the higher dimension spectrum. In particular we see that a sufficient condition for $j_0=1$ is that that the average anomalous dimensions are order one.

Before moving on, we should be more precise about what we mean by $\simeq$ above, since alas elegance comes at the cost of rigor. The convergence (divergence) of the original integral actually only leads to constraints on the limit inferior (superior) of $\overline{\gamma_n^2}$:
\bea
\overline{\gamma_n^2}\underset{n\to \infty}{\simeq}  \Delta_n^{-2(1-\tj)} \qquad \equiv \qquad  
\left\{
\begin{array}{cc}
\liminf_{n\to \infty} \overline{\gamma_n^2} \,\Delta_n^{2(1-\tj)+\underline{\epsilon}}<\infty\,, & \underline{\epsilon}\geq 0\\
& \\
\limsup_{n\to \infty} \overline{\gamma_n^2} \, \Delta_n^{2(1-\tj)-\overline{\epsilon}}<\infty\,, & \overline{\epsilon}\geq 0
\end{array}
\right.
\eea
What we expect then is that a suitable choice of $C_n$ in the definition of $\overline{\gamma_n^2}$ should lead to the two parameters $\overline{\epsilon},\underline{\epsilon}$ being zero. In practice we expect that $C_n$ is likely an $n$ independent, O(1) constant. This is because the OPE density is strongly constrained inside unit size bins, both from above and below, as are anomalous dimensions of operators inside such bins \cite{Mazac2019a}. We leave a more detailed understanding of this for future work. 

\subsection{General bounds}

A general way to constrain $\tj$ is to use the positivity properties of the double discontinuity. Choose any subset $\mathcal S$ of non-identity operators appearing in the correlator. We have:
\bea
0&\leq \, (1-z)^{2\Df} \mbox{dDisc}_s \left[\mathcal G(z,z)-G_{0,0}(z,z|\Df)-\sum_{\mathcal S} a_{\Delta,\ell} G_{\Delta,\ell}(z,z|\Df)\right]\\
&\underset{z\to 1}{\sim}\nu\, (1-z)^{1-\tj} -2 (1-z)^{2\Df}\left[\sin^2(\pi \Df) +\sum_{\mathcal S} a_{\Delta,\ell} \sin^2\left[\frac{\pi}2(\Delta-2\Df)\right] c_{\Delta,\ell} g(1-z)\right]
\eea
where the constants $c_{\Delta,\ell}>0$ are irrelevant and $g(1-z)$ captures the $\bar z=z\to 1$ behaviour of conformal blocks:
\bea
g(1-z)\underset{z\to 1}= \left\{
\begin{array}{cl}
O[\log(1-z)] & d=1 \\
O[\log(1-z)^2] & d=2 \\
O[(1-z)^{1-\frac{d}2}] & d>2\,, d\neq 4,6,\ldots\\
O[\log(1-z)(1-z)^{1-\frac{d}2}] & d=4,6,8\ldots
\end{array}
\right.
\eea
Choosing larger sets $\mathcal S$ for which we can resum contributions leads to constraints on $\tj$. The simplest bound is found by setting $\mathcal S$ to consist of a single term. In this case we find
\bea
\tj\geq \mbox{max}\left\{1,\frac d2\right\}- 2\Df\,.
\eea
More precisely, the stronger bound applies as long as the correlator contains at least one non-identity operator which is not annihilated by the double discontinuity, i.e. if the correlator is not that of generalized free fields. 

We can do better by considering a larger set of states, and one way to do this is as follows. The lightcone bootstrap \cite{Komargodski2013,Fitzpatrick2013,Caron-Huot:2017vep} tells us there must be towers of states at large spin whose OPE coefficients are approximately those of a generalized free field, and whose anomalous dimensions depend on the leading twist operator in the correlator. Let us call the set of such states $\mathcal S_0$. The lightcone bootstrap tells us
\bea
\lim_{z\to 1} \lim_{\bar z\to 0} \sum_{\mathcal S_0} 
\gamma_{0,\ell} a_{\Delta,\ell}^{\mbox{\tiny free}} G_{\Delta,\ell}(z,\bar z|\Df)\sim (1-z)^{\frac{\tau_0}2-\Df}\,.
\eea
where the anomalous dimensions take the form
\bea
\gamma_{0,\ell}\equiv \Delta-2\Df-\ell \underset{\ell\to \infty}\sim \frac{\gamma_0}{\ell^{\tau_0}}
\eea
for $d>2$. Examining our expression for the double discontintuiy, we see that we are interested in a different, but still calculable limit:
\bea
\lim_{z\to 1} \sum_{\mathcal S_0} 
\gamma_{0,\ell}^2 a_{\Delta,\ell}^{\mbox{\tiny free}} G_{\Delta,\ell}(z,z|\Df)\sim (1-z)^{-\mbox{\tiny max}(\Df-\tau_0,0)-\frac{d-2}2}\,,
\eea
This leads to a better bound for $d>2$:
\bea
\tj>\frac d2 - 2\Df +\mbox{max}(\Df-\tau_0,0)\,.
\eea
A simple cross-check is to consider generalized free correlators of non-elementary fields, for which we have
\bea
\mathcal G(z)-\mathcal G^B(z) \supset \frac{1}{z^{2\Df-\alpha \Df}(1-z)^{\alpha \Df}} +\ldots+ \frac{1}{(1-z)^{\alpha \Df}(1-z)^{2\Df-\alpha\Df}} ,\qquad \alpha\in(0,1)
\eea
A simple computation shows that in this case $\tj =1-\tau_0$ with $\tau_0=\alpha \Df$. Our bound is then satisfied as a consequence of the unitarity bound $\Df\geq \frac{d-2}2$\,.

These bounds are rather modest except for rather small values of $\Df$. For instance, for the spin-field correlator in the 3d Ising model $\tau_0=1$ and hence in this case one finds
\bea
j_0^{\sigma}> \frac 12 -2\gamma_\sigma\sim 0.464\,,
\eea
which is compatible with the recent estimate $J_0\sim 0.8$ appearing in \cite{Caron-Huot:2020ouj}.
This should be contrasted with the same bound as applied to the energy operator four-point function,
\bea
j_0^{\epsilon}>\frac{3}2-\Delta_{\epsilon}-1\sim -0.91\,.
\eea
In fact, below we will improve this to $j_0^{\epsilon}\geq 1-\Delta_{\epsilon}\sim -0.41$. Note that while $J_0$ should be the same for both correlators, this is not necessarily the case for $j_0$.
\subsection{Dispersion relation analysis}

Let us now examine what we can learn about the Regge limit from a 1d perspective, i.e. from knowledge of the correlator restricted to the line. At first it might seem puzzling how we can ever study the Regge limit once we've set $z=\bar z$, since we then cannot separately continue $z$ and $\bar z$. But as explained in \cite{Mazac:2018} and as we now review, this is not a problem for accessing the $u$-channel Regge limit. As usual, let us set here $\mathcal G(z)$ to be that analytic function which matches $\mathcal G(z,z)$ in the range $z\in(0,1)$. We notice that
\bea
\mathcal G(z)\underset{\mbox{\tiny Im}\, z>0}{=}(1-z)^{-2\Df}\sum_{\Delta} e^{i\pi (\Delta-2\Df)} a_{\Delta} \frac{G_{\Delta}(\mbox{$\frac z{z-1}$})}{(\mbox{$\frac z{z-1}$})^{2\Df}}
\eea
It follows that taking $z$ to infinity in the upper half-plane is equivalent to the Regge limit:
\bea
\frac{\left[\mathcal G(z)-(1-z)^{-2\Df}\mathcal G(\mbox{$\frac{z}{z-1}$})\right]}{i}\underset{z\to \infty}{\sim}\frac{\mu}{(-z)^{1-j_0}}\,, \qquad \mbox{Im}\,z>0\,.
\eea
If we insist on writing the limit in the $t$-channel we have equivalently,
\bea
\frac{\left[\mathcal G(\mbox{$\frac{z}{z-1}$})-(1-z)^{2\Df}\mathcal G(z)\right]}{i} \underset{z\to 1}{\sim}\mu (1-z)^{1-j_0}\,, \qquad \mbox{Im}\, z<0\,.
\eea

Our first task is to better understand the relationship between the Regge limit of the full correlator versus that of its double discontinuity. Our tool is the (bosonic) dispersion relation \reef{eq:dispersions} for the subtracted correlator:\footnote{Had we defined the Regge limit by adding the Euclidean correlator (instead of subtracting it) we would have used instead the fermionic dispersion relation.}
\bea
\overline{\mathcal G}(w)&=-\int_0^1\ud z \, \ghu_w(z|\Df)\, \mbox{dDisc}_B\left[\overline{\mathcal G}(z)\right]\,.
\eea
The definition of $\tj$ implies
\bea
(1-z)^{2\Df} \mbox{dDisc}_B\, \overline{\mathcal G}(z)\underset{z\to 1}\sim\nu (1-z)^{1-\tj}\,.
\eea
We would like to understand what this implies for the behaviour of the correlator in the limit of large $w$. Naively this is determined by expanding $\ghu_w$, but in reality we must be careful since in general the limit might not commute with the integration. Indeed we have
\bea
\ghu_w(z|\Df)\underset{z\to 1, w\to \infty}\sim\frac{2}\pi \frac{i \,w}{(w-\frac 1{1-z})(w+\frac{1}{1-z})}\, (1-z)^{2\Df-2}\,, \qquad  \quad \mbox{for fixed}\quad (1-z)w\,.\label{eq:glim}
\eea
We have found this expression in every kernel we have examined. More generally it follows from the functional equation close to the delta function singularities. We see that naively taking the large $w$ limit in $g_w(z)$ will lead to a divergent integration region close to $z=1$ unless $\tj<0$. Therefore in general we {\em cannot} take the limit inside integral and the large $w$ behaviour depends on the precise range of $\tj$. We will therefore split our analysis into two cases, beginning with the more interesting one $\tj>0$.

\subsubsection{$\tj>0$}

In this case we can use \reef{eq:glim} to find:
\bea
\overline{\mathcal G}(w)\underset{w \to \infty}{\sim}-\frac{\nu \,\kappa(1-\tj)}{(-w)^{1-\tj}}\,, \qquad \kappa(x):=\frac{e^{-i \frac{\pi}2 x}}{\cos\left(\frac{\pi}2 x\right)}\,.
\eea
Recall the definition of $\overline{\mathcal G}$,
\bea
\overline{\mathcal G}(w)=\mathcal G(w)-\mathcal G^B(w)-\sum_{0<\Delta\leq 2\Df} a_{\Delta} \mathcal P_{\Delta}(z)\,.
\eea
To proceed we need to understand the asymptotics of the Polyakov blocks and this is done in appendix \ref{app:asymp}. There we show that:
\bea
\mathcal P_{\Delta}(w) \underset{z\to \infty}\sim \frac{\kappa(\Delta)}{(-w)^{\Delta}}+O(1/w)
\eea
From this we determine the large $w$ behaviour of the full correlator:
\begin{eqnarray}
\label{eq:largew}
\boxed{
\tj>0:\qquad\mathcal G(w)\underset{w \to \infty}{\sim}1+a_{\Delta_g} \frac{ \kappa(\Delta_g)}{(-w)^{\Delta_g}}-
\frac{\nu \,\kappa(1-\tj)}{(-w)^{1-\tj}}+\ldots
}
\end{eqnarray}
This equation is one of the main result of this section. It relates the large $w$ behaviour of the full CFT correlator in terms of the Regge limit of its double discontinuity. We see that it is made up of two sets of terms, a set of  essentially trivial contributions which should be thought of as arising from the $u$-channel OPE, and a non-trivial contribution coming from the double-discontinuity. Since $\tj>0$ by assumption, the contribution from the lowest dimension operator is only important if $\Delta_g<1$. This expression is certainly only correct up to terms of order $w^{-1}$, although there could also be subleading terms coming from the double discontinuity which are even more important. 
 In fact such subleading terms have the possibility to become the dominant ones in case of cancellations between the trivial and non-trivial pieces above, which is allowed by the crucial minus sign in front of the double discontinuity term. This result has applications to the study of the out-of-time-order correlator, as we will mention in the discussion.

The Regge limit of the correlator is now determined by the result given above:
\begin{multline}
\frac{\left[\mathcal G(w)-(1-w)^{-2\Df}\mathcal G(\mbox{$\frac{w}{w-1}$})\right]}i\underset{w\to \infty}{\sim} \frac{i \nu\,\kappa(1-\tj)}{(-w)^{1-\tj}}-a_{\Delta_g} \frac{\tan\left(\frac{\pi \Delta_g}2 \right)}{(-w)^{\Delta_g}}+
\ldots
\end{multline}
Unlike before, it is now manifestly impossible to have cancellations between trivial and non-trivial pieces. In fact the definition of the Regge limit tells us 
\begin{eqnarray}
\boxed{
j_0=\mbox{max}\{1-\Delta_g,\tj\} \quad \Rightarrow \quad j_0 \geq 1-\Delta_g\quad \mbox{and}\quad \left\{\begin{array}{cc} \mu=-a_{\Delta_g} \tan\left(\frac{\pi \Delta_g}2\right) & j_0>\tj\\
\mu=i \nu\, \kappa(1-\tj) & j_0=\tj
\end{array}
\right.
}
\end{eqnarray}
This result is perfectly consistent with the relation between $j_0$ and $\tj$ as described in \reef{eq:relj0}. The dispersion relation therefore tells us that either $j_0=\tj$ or if $j_0 > \tj$ then it is determined in terms by low energy data corresponding to ``subtractions''.

\subsubsection{$-1<\tj\leq 0$}

Let us now discuss the more involved case where $\tj\leq 0$. To be thorough we would have had to split the analysis into several regions\,,
\bea
-k \geq \tj \geq -k-1\,, k \in \mathbb Z_{\geq 0}\,,
\eea
and study subleading contributions to the large $w$ limit of Polyakov blocks. Here we will focus on the case $k=0$ so that $-1<\tj\leq 0$.

In this case, an examination of the dispersion relation and the functional kernel shows us that there are now both analytic and non-analytic contributions in the large $w$ limit.
\bea
\overline{\mathcal G}(w)\underset{w \to \infty}{\sim}-\frac{\nu \,\kappa(1-\tj)}{(-w)^{1-\tj}}-i \pi\,\frac{\sum_{\Delta\geq 2\Df} a_{\Delta}\tilde \beta_{0}(\Delta)}{w}\,.
\eea
As explained in appendix \ref{app:asymp}, the $\tilde \beta_{0}$ is a functional bootstrapping correlators decaying faster than $z^{-1}$ at infinity. This follows from the asymptotic expression of Polyakov blocks:
\bea
\mathcal P_{\Delta}(w) \underset{w\to \infty}\sim \frac{\kappa(\Delta)}{(-w)^{\Delta}}-i \pi\,\frac{\tilde \beta_{0}(\Delta)}{w}\,.
\eea
Hence the Regge limit of the correlator is now:
\begin{multline}
\frac{\left[\mathcal G(w)-(1-w)^{-2\Df}\mathcal G(\mbox{$\frac{w}{w-1}$})\right]}i\underset{w\to \infty}{\sim} 
\frac{i\nu\,\kappa(1-\tj)}{(-w)^{1-\tj}}\\
-a_{\Delta_g} \frac{- \tan\left(\frac{\pi \Delta_g}2 \right)}{(-w)^{\Delta_g}}-\pi\,\frac{\sum_{\Delta\geq \Delta_g} a_{\Delta}\tilde \beta_{0}(\Delta)}{w}+\ldots
\end{multline}
Let us discuss what this expression implies for $j_0$. The analysis depends on the gap:
\bea
\Delta_g<1:& \qquad j_0=1-\Delta_g\\
\Delta_g>1:& \qquad j_0=0 \quad \mbox{or}\quad j_0=\mbox{max}\{1-\Delta_g,\tj\}
\eea
The case $\Delta_g=1$ is special: the divergence in $\tan(\pi \Delta/2)$ cancels against a divergence in $\tilde \beta_0(\Delta)$ as $\Delta\to 1$ to give an asymptotic behaviour of the form $\log(w)/w$. When $\Delta_g>1$, we can in principle have a term of the form
\bea
\mathcal G(w) \underset{w\to \infty}  \supset  -i\pi\frac{\sum_{\Delta>1} a_{\Delta} \tilde \beta_0(\Delta)}{w}
\eea
which would imply $j_0=0$. This is however quite strange. Naively,
\bea
\sum_{\Delta>1} a_{\Delta} \tilde \beta_0(\Delta)\overset{?}{=}\tilde \beta_0\left[ \mathcal G(w)-\mathcal P_0(w)\right]-(w\leftrightarrow 1-w)=0
\eea
but we've just seen that the large $w$ behaviour of $\mathcal G-\mathcal P_0$  can be $O(1/w)$ and so this manipulation is not allowed since $\tilde \beta_0$ is only crossing compatible for correlators decaying {\em faster} than $1/w$ at infinity.  So this $j_0=0$ case would correspond to a strange situation where even though the sum rule of $\tilde \beta_0$ converges, the action of the functional does not commute with the full sum over states. Although strange, this is not impossible. Suppose for instance that the correlator $\mathcal G(w)$ was a finite sum of Polyakov blocks. For instance:
\bea
\mathcal G(w)-\mathcal P_0(w)=a_g \mathcal P_{\Delta_g}(w) \underset{w\to \infty}{\sim} -i\pi \frac{a_g \tilde \beta_0(\Delta_g)}{w}
\eea
with $\Delta_g>1$. This case would indeed give $j=0$, and is perfectly consistent with the above. The $\tilde \beta_0$ sum rule trivially converges to a finite result, simply because even though the full correlator has an infinite number of conformal blocks, there are only a finite subset of them where $\tilde \beta_0$ is actually non-zero. This example is of course contrived, since this is not a physical correlator, in particular it does not have a unitary OPE. But our dispersion relation must allow for such cases too, which is perhaps an explanation for the strange possibility of $j_0=0$ above. It is natural to conjecture that for physical unitary correlators this is impossible, but we were unable to prove it.

In any case, this analysis tells us something more interesting, which is that if $j_0<0$ then necessarily the $\tilde \beta_0$ sum rule must vanish, and this has consequences for the spectrum:
\begin{eqnarray}
\boxed{
-1<\tj<0: \qquad j<0 \quad \Leftrightarrow \quad \Delta_g \in (1,2\Df)\quad \mbox{and}\quad \sum_{\Delta>\Delta_g} a_{\Delta} \tilde \beta_0=0
}
\end{eqnarray}
To conclude let us briefly comment on the generalization to other $k$. As we allow $
\tj$ to become more negative we must take into account more and more subleading terms in the large $w$ expansion of Polyakov blocks. Each subleading asymptotics should be controlled by a functional. Hence we expect that demanding $j_0$ more and more negative will impose more and more constraints on the spectrum. 

\section{Application: 3d Ising model on the line}
\label{sec:approxIsing}
Any higher dimensional CFT four-point correlator becomes a 1d CFT correlator on the line $z=\bar z$, this means that any such correlator must satisfy the Polyakov bootstrap constraints:
\bea
\mathcal G(z,z)=\sum_{\Delta} a_{\Delta} G_{\Delta}(z|\Df)=\sum_{\Delta} a_{\Delta} \mathcal P_{\Delta}(z|\Df)\,.
\eea
Polyakov blocks are roughly suppressed by the squared anomalous dimension of operators, which means that CFT correlators which approach generalized free fields in the UV can be efficiently represented as finite sums of Polyakov blocks. We will use this fact to obtain a very accurate representation of the 3d Ising model spin field correlator on the line. To do this we need an efficient and accurate method for computing Polyakov blocks and this is provided by the dispersion relations \reef{eq:polyposrep}.

It is natural to organize the computation in terms of 3d conformal group primaries. To do this we begin by noting that 3d conformal blocks on the line are actually known analytically \cite{Rychkov:2015lca}:
\bea
G_{\Delta,\ell}^{3d}(z|\Df)=z^{-2\Df}\frac{(4\rho)^{\Delta}}{1-\rho(z)^2}\, _4F_3\left(\begin{array}{cccc}
\frac 12 &\frac{\Delta-\ell-1}2&\frac{\Delta+\ell}2& \Delta-1\\ &\Delta-\frac 12&\frac{\Delta+\ell+1}2&\frac{\Delta-\ell}2\end{array}; \rho(z)^2\right)
\eea
with $\rho(z)=(1-\sqrt{1-z})/(1+\sqrt{1-z})$. This allows us to define a resummed Polyakov block which captures the contributions of all $SL(2,\mathbb R)$ primaries contained in $G_{\Delta,\ell}^{d=3}$:
\bea
\mathcal P_{\Delta,\ell}^{3d}(w|\Df):=G_{\Delta,\ell}^{3d}(w|\Df)-\OU_w\left[F_{\Delta,\ell}^{3d}\right]
\eea
where $F_{\Delta,\ell}^{3d}$ is the 3d crossing vector defined analogously to the $d=1$ case. We can compute the resummed block via the dispersion formula:
\bea
\mathcal P_{\Delta,\ell}^{3d}(w)=-2\sin^2\left[\frac{\pi}2(\Delta-2\Df)\right]\,\int_0^1 \ud z\, \gu_w(z)\, G_{\Delta,\ell}^{3d}(z|\Df)\,,\qquad \Delta \geq 2\Df
\eea
We should point out that the identity contribution trivially remains the same:
\bea
\mathcal P^{3d}_{0,0}(z)=\mathcal P_0(z)=\mathcal G^B(z)=1+z^{-2\Df}+(1-z)^{-2\Df}\,.
\eea 
The Polyakov bootstrap constraints become
\bea
\mathcal G^{3d}(z,z)=\sum_{\Delta,\ell} a_{\Delta,\ell} \mathcal P_{\Delta,\ell}^{3d}(z)
\eea

Let us apply this representation to the 3d Ising model correlator of the spin operator $\sigma$, $\mathcal G_{\mbox{\tiny Ising}}(z)\sim \langle \sigma \sigma \sigma \sigma\rangle$, setting $\Df=\Delta_{\sigma}$. 
The extremal functional method allows for accurate numerical determinations of the 3d Ising spectrum \cite{ElShowk:2012hu,ElShowk:2012ht,El-Showk:2014dwa,Simmons-Duffin:2016wlq}. The low lying operators in the $\sigma$ correlator are described in table \ref{tab:ising}. More generally one finds that the spectrum is neatly organized into Regge trajectories of operators. In particular one finds the smallest twist trajectory very rapidly approaches $\Delta(J)\sim 2\Delta_{\sigma}+2n+J$. This means that the contribution of this trajectory will be suppressed when computing the double discontinuity. 
\begin{table}[h]
\begin{center}
\begin{tabular}{|c|c|c|c|}
\hline
$\mathcal O$ & $\Delta_{\mathcal O}$ & $\ell$ & $a_{\mathcal O}\equiv a_{\Delta,\ell}$\\
\hline $\sigma$ & 0.5181489 & 0 & - \\
\hline $ \epsilon$ & 1.412625 & 0 & 1.1064\\
\hline $ \epsilon'$ & 3.82968 & 0 & 0.00281\\
\hline $ T $ & 3 & 2 & 0.2836 \\
\hline $ C $ & 5.0227 & 4 & 0.01745\\ \hline
\end{tabular}
\end{center}
\caption{Low lying operators in the 3d Ising model. The OPE data refers to couplings in the spin-field four-point correlator, i.e. $a_{\Delta,\ell}=\lambda^2_{\sigma \sigma \mathcal O_{\Delta,\ell}}$. Data taken from \cite{Simmons-Duffin:2016wlq}.}
\label{tab:ising}
\end{table}

The first thing we can check is that the dispersion relation actually holds,
\bea
\mathcal G_{\mbox{\tiny Ising}}(w)=\mathcal G^B(w)-\int_0^1 \ud z \,\gu_w(z) \mbox{dDisc}_B\left[\sum_{\Delta>0,\ell} a_{\Delta,\ell} G_{\Delta,\ell}^{3d}(z|\Df)\right]\,.
\eea
In order to check the validity of this formula, as well as for computing the Polyakov blocks, we need to compute $\gu_w(z)$ for $\Df=\Delta_{\sigma}$. We do this by numerically solving the Fredholm equation \reef{eq:dispf}.\footnote{As explained in that section, the equation for $\fu_w(z)$ does not have a unique solution. However in practice a numerical solution, obtained by discretizing and solving the equation as a linear operator acting on $\fu_w$, always gives us a well defined unique solution.} Plugging in the numerical spectrum and normalizing by the identity contribution we reassuringly find the dispersion relation is satisfied to about $10^{-5}$.

\begin{figure}%
\begin{center}
\hspace{-2 cm}
\includegraphics[width=12 cm]{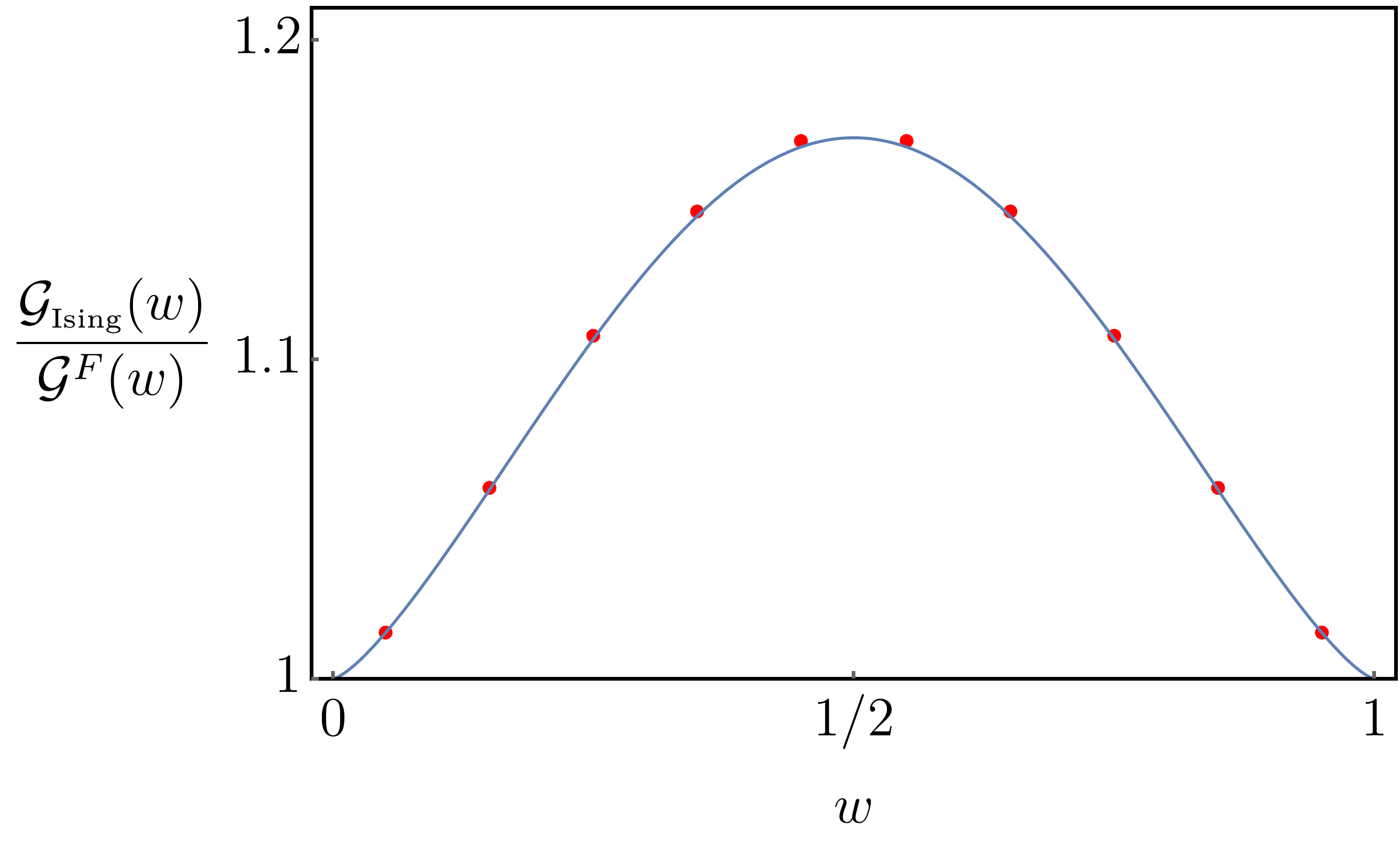}%
\caption{Comparison between the 3d Ising spin field correlator and its approximation as a sum of two Polyakov blocks. The results are normalized relative to the generalized free fermion correlator $\mathcal G^F(w)$. The line is the correlator as computed from the conformal block expansion up to dimension 10, plotted in the interval $w\in(0,1/2)$ which is then mirrored to get a crossing symmetric shape. This is not necessary for the Polyakov block approximation which is automatically crossing symmetric and shown as the red dots.}%
\label{fig:comparison1}%
\end{center}
\end{figure}

The point now is that this result is almost unchanged if, instead of all low lying states, we include the contribution of a single term in the double discontinuity, namely the the energy operator $\mathcal \epsilon$. This is due to a combination of operators having small anomalous dimensions and rapid decay of OPE coefficients at large dimension. What this means in terms of the Polyakov bootstrap is that we have the following approximation for the Ising correlator:
$$
\boxed{
\mathcal G_{\mbox{\tiny Ising}}(z)= \mathcal G^B(z)+a_{\mathcal \epsilon} \mathcal P_{\Delta_{\epsilon},0}^{d=3}(z) \quad \pm 0.1\%
}
$$
where the error holds for $z\in(0,1)$. A comparison between the correlator and its approximation is shown in figure \ref{fig:comparison1}.

We find it amusing that such a good approximation to the correlator is obtainable with just two Polyakov blocks and three parameters, $\Delta_\sigma$, $\Delta_\epsilon$ and $a_{\epsilon}$. In fact, it is possible to get an almost equally good approximation that relies on just two parameters. Essentially this is because in the above representation the OPE coefficient of $\epsilon$ is not really arbitrary, since the contribution from the $n=0$ operator in the Polyakov blocks (of dimension $\Delta=2\Delta_\sigma$) must (approximately) drop out of the OPE. This approximately fixes $a_{\epsilon}$:
\bea
a_{\epsilon} \sim 2/ \alpha^B_{0}[F_{\Delta_{\epsilon},0}^{d=3}]\,.
\eea
Similarly it may seem odd that the contribution of the stress tensor $T$ need not be included, but this can be understood from the smallness of its ``anomalous dimension'', $\Delta_T-2\Delta_{\sigma}-2\sim -0.036$. The contribution of $T$ is then in fact approximately reproduced from the $n=1$ double trace operator contained in the Polyakov blocks. For instance, the OPE coupling to $T$ becomes:
\bea
a^{\mbox{\tiny free}}_{\Delta^B_{1}}-a_{\epsilon} \alpha^B_{1}[F_{\Delta_{\epsilon},0}^{d=3}]\sim 0.286 \sim a_T\,.
\eea

To conclude this section, we point out that a different way of thinking about all this is that the 3d Ising correlator on the line is very well approximated by a {\em purely} 1d solution to crossing (i.e. not exactly representable in terms of higher-D conformal blocks), namely that solution which maximizes the OPE coefficient of the operator of dimension $\Delta_{\epsilon}$. We work this out in more detail in the next section.

\subsection{Ising correlator and interacting Polyakov blocks}
\label{sec:interactingpoly}

In this paper we have derived two dispersion relations, both associated to ``free'' solutions to crossing, one for a boson another for a fermion. We used the bosonic one to describe the 3d Ising correlator. It would not have been a very good idea to use the fermionic one, since the fermionic double discontinuity would not be small. More generally, we expect that these basis are just two of an infinite set of functional bases, and hence we may wonder if there is an even better basis with which to describe the correlator.

We expect that for any 1d functional basis we can construct an associated dispersion relation, and not just those which are associated to the generalized free solutions to crossing. To each such basis there will be a version of the Polyakov bootstrap with ``interacting'' Polyakov blocks which have the property that they will vanish quadratically whenever the dimension $\Delta$ is tuned to one of the dimensions $\Delta_n^*$ in the associated extremal solution.

Concretely we have in mind a set of functionals satisfying relations analogous to those of the bosonic basis:
\bea
\alpha_n^*(\Delta_m^*)&=\delta_{n,m}\,,& \qquad \partial_{\Delta}\alpha_n^*(\Delta_m^*)&=-c_n^* \delta_{n,0}\,,\\
\beta_n^*(\Delta_m^*)&=0\,,& \qquad \partial_{\Delta}\beta_n^*(\Delta_m^*)&=\delta_{nm}-d_n^* \delta_{n,0}\,
\eea
for some carefully chosen spectrum $\Delta_n^*$ corresponding to an extremal solution to crossing. That is, such that the following equation holds:
\bea
\sum_{n=0}^\infty a_n^* F_{\Delta_n^*}(z)=-F_0(z)\,,\qquad a_n^*=-\alpha_n^*(0)\,, \qquad \beta_n^*(0)=0\,.
\eea
With such functionals in hand we can construct the associated Polyakov blocks,
\bea
\mathcal P_{\Delta}^*(z)=G_{\Delta}(z|\Df)-\sum_{n=0}^{\infty}\left[ \alpha_n^*(\Delta) G_{\Delta_n^*}(z|\Df)+\beta_n^*(\Delta) \partial_{\Delta} G_{\Delta_n^*}(z|\Df)\right]\,.
\eea
Note that in particular that Polyakov blocks will have double zeros on the spectrum $\Delta_n^*$ and that the identity Polyakov block computes the extremal solution:
\bea
\mathcal P_0^*(z)=\mathcal G^*(z)=G_0(\Df)+\sum_{n=0}^\infty a_n^* G_{\Delta_n^*}(z)\,.
\eea
The crossing constraints would then be equivalent to an ``interacting'' version of the Polyakov bootstrap:
\bea
\sum_{\Delta}a_{\Delta} F_{\Delta}(z)=0 \Leftrightarrow \mathcal G(z)=\sum_{\Delta} a_{\Delta} \mathcal P_{\Delta}^*(z|\Df)
\eea
We could equally well define the associated master functional:
\bea
\Omega^*_w:=\sum_{n=0}^\infty\left[\alpha_n^* G_{\Delta_n^*}(w)+\beta_n^* \partial_{\Delta} G_{\Delta_n^*}(w)\right]
\eea
the action of which on the crossing equation would lead to the above relation, or equivalently to an ``interacting'' dispersion relation for the correlator.

From this discussion, the conclusion is that if we could find an extremal solution whose spectrum is given by $\Delta_\epsilon$ together double trace operators with small anomalous dimensions, then we could essentially replace our approximation to the Ising correlator by a single term, namely the identity Polyakov block of this extremal solution. As it turns out there is indeed one such solution: the correlator which maximizes the OPE of an operator of dimension $\Delta_{\epsilon}$. Let us therefore henceforth define the starred quantities as pertaining to this particular solution. We expect:
$$
\boxed{
\mathcal G_{\mbox{\tiny Ising}}(z)\simeq \mathcal P_0^*(z)=\mathcal G_{\mbox{\tiny opemax}}(z|\Delta_\sigma; \Delta_{\epsilon})
}
$$
To reiterate, we expect this to be true since other Polyakov blocks should be suppressed, either by their OPE coefficient or by the fact that their dimensions lie close to some $\Delta_n^*$.

The discussion above might have seemed somewhat abstract and containing a lot of what ifs. Nevertheless, it is possible to make it more precise in a systematic way. What we are interested in is to determine an approximation to $\mathcal P_0^*$. While determining the fully interacting functional basis is difficult, we can obtain very good approximations by expressing everything in terms of the bosonic functional basis over which we have good control. In turn this works because in the fully interacting solutions the dimensions $\Delta_n^*$ rapidly asymptote to the free bosonic ones $\Delta_n^B$ as $n$ increases. This means that expressing interacting functionals in the bosonic basis requires only a handful of terms.

In practice we could determine both the extremal solution and associated functionals by simply running the usual numerical bootstrap algorithm for the OPE maximization problem using a truncated set of bosonic functionals \cite{Paulos:2019fkw}. Alternatively we could have used the same method to approximately solve the correlator maximization problem subject to a gap $\Delta_g=\Delta_\epsilon$, since the associated functional is nothing but $\Omega^*$ whose action on the identity computes $\mathcal P_0^*$.
For our purposes it will actually be sufficient to work in an approximation where we truncate the basis to $\alpha_{0,1}^B$ and $\beta_1^B$. In this case we can solve the optimization problem ``by hand''. 

The interacting functionals are defined by:
\bea
\alpha_{n}^*=\alpha_n^B-a_{n0} \alpha_0^B-a_{n1} \alpha_1^B-b_{n0} \beta^B_1 \,,\\
\beta_{n}^*=\beta_n^B-c_{n0} \alpha_0^B-c_{n1} \alpha_1^B-d_{n0} \beta^B_1 \,,
\eea
We begin by imposing:
\bea
\beta^*_1(\Delta_{\epsilon})=0,\qquad \beta^*_1(0)=0\,
\eea
After doing this $\beta^*_1(\Delta)$ now has a zero at $\Delta_1^*=2\Df+2+\gamma_2$, where $\gamma_2$ turns out to be $\sim 0.0377$.  The remaining constant can be fixed by demanding $\partial_{\Delta} \beta_1^*(\Delta_1^*)=1$. Similarly we impose
\bea
\alpha_0^*(\Delta_\epsilon)=1\,,\quad \alpha_0^*(\Delta_1^*)=0\,,\quad \partial_{\Delta} \alpha_0^*(\Delta_1^*)=0\,,\\
\alpha_1^*(\Delta_\epsilon)=0\,,\quad \alpha_1^*(\Delta_1^*)=1\,, \quad\partial_{\Delta} \alpha_1^*(\Delta_1^*)=0
\eea
which fixes $\alpha_0^*$ and $\alpha_1^*$. In this approximation, no other operator picks up anomalous dimensions so that $\Delta_n^*=\Delta^B_n$ for $n\geq 2$, so that the remaining functionals can then be determined by imposing that they have double zeros on $\Delta_1^*$ and a simple zero $\Delta_\epsilon$.

\begin{figure}[t!]%
\begin{center}
\hspace{-2 cm}
\includegraphics[width=12 cm]{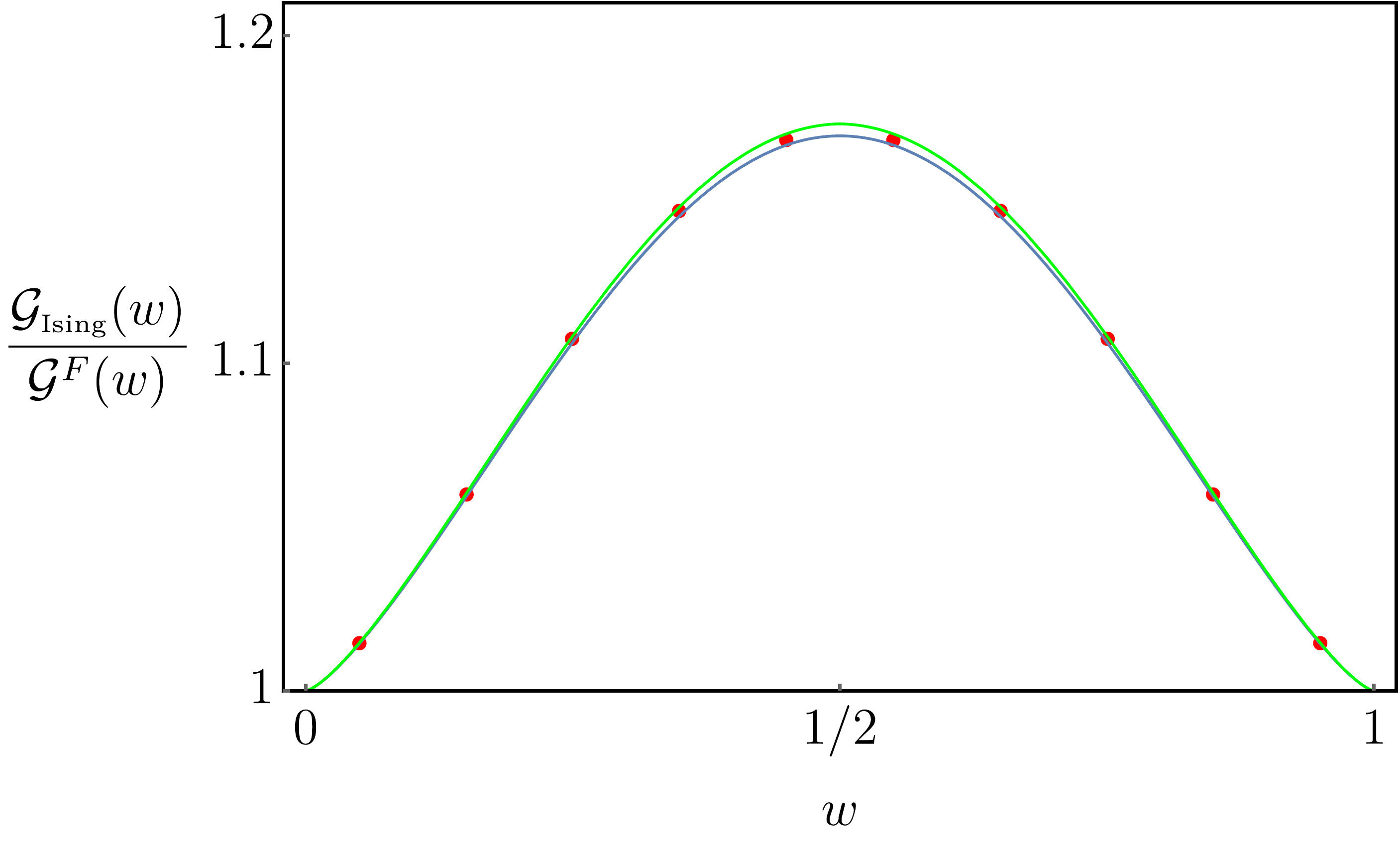}%
\caption{Comparison between the 3d Ising spin field correlator and its approximations. The results are normalized relative to the generalized free fermion correlator $\mathcal G^F(w)$. The blue line (the lowest) is the correlator as computed from the conformal block expansion up to dimension 10. The red dots represent the two Polyakov block approximation of section \ref{sec:approxIsing}. The green line is the identity interacting Polyakov block described in the text, computed from its conformal block expansion.}%
\label{fig:comparison2}%
\end{center}
\end{figure}

The OPE coefficient of the operator of dimension $\Delta_\epsilon$ is determined by
\bea
a_{0}^*=-\alpha_0^*(0)=1.1211\,,
\eea
and is a rigorous upper bound on the OPE coefficient of such an operator in any solution to crossing. Notice this is very close to the 3d Ising result $a_\epsilon=1.106$. Using a larger basis would lead to an even better match.

Our tentative extremal solution to crossing is 
\bea
\sum_{n=0}^{\infty} a_n^* F_{\Delta^*_n}(z)\overset{?}{=} -F_0(z)\,,\qquad a_n^*=-\alpha_n^*(0)
\eea
Applying our functional basis to the equation above we find that all the $\alpha_n^*$ constraints are satisfied with our definition of the OPE coefficients. However the equations involving $\beta_n^*$ with $n\geq 2$ fail, since $\beta_n^*(0)$ will be small but not zero.
By increasing the size of the basis we can obtain more and more accurate anomalous dimensions and OPE coefficients, and the functional sum rules will be better and better satisfied.

In figure \ref{fig:comparison2} we plot $\mathcal G^*(z)$ in this approximation against the 3d Ising correlator. We see that both expressions indeed closely match (to about $0.5\%$). It should be possible to get an improvement by computing $\mathcal P_0^*$ with more accuracy, but we draw the line here. A nice way to understand this result is to notice that the 3d Ising correlator is such that it maximizes the OPE coefficient of the operator $\Delta_{\epsilon}$ for all 3d CFTs. This is a triviality, in that setting the gap to be $\Delta_{\epsilon}$ the solution is unique. What is surprising then is that this still remains approximately true when we include only those constraints arising from an $SL(2,\mathbb R)$ subset of the conformal group.

\section{Conclusions}

\label{sec:discussion}
\subsection{Comments on higher dimensions}

We will now discuss a few similarities and differences between
what we have presented in this paper and analogous works in higher dimensions.

In \cite{Carmi:2019cub} a dispersion relation was written down for CFT correlators in arbitrary dimensions, starting from the Lorentzian inversion formula \cite{Caron-Huot:2017vep}. This dispersion relation was subsequently related to a certain set of extremal functionals in \cite{Mazac:2019shk,Caron-Huot:2020adz}. The logic is very similar to what we've done in this work: there are master functionals which act as generating functions for an extremal functional basis which is associated to the generalized free field solution to crossing. This master functional leads to a sum rule which can be written as a CFT dispersion relation. The dispersion relation can be alternatively interpreted as encoding a version of the Polyakov bootstrap in higher dimensions in terms of so-called Polyakov-Regge blocks, whose OPE decomposition encodes the functional actions of the basis.

We should point out however that completeness of the higher dimensional functional basis has not yet been rigorously established. This means that, at this point, it is not completely clear that the constraints encoded implicitly in the dispersion relation of \cite{Carmi:2019cub} are the same or stronger than those implied by the the functionals constructed in \cite{Mazac:2019shk}. At least one way to think about this possible difference is to think about the singularity structure of CFT correlators. Indeed talking about completeness of a set of functionals only makes sense once we restrict ourselves to classes of correlators satisfying appropriate asymptotic behaviours. But there are yet kinematical regimes where our understanding of correlators is lacking. In particular, functionals acting on the first sheet in cross-ratio space probe the double lightcone limit of CFT correlators, a limit which is not yet understood. Depending on the properties of this limit, the set of allowed functionals can be more or less restricted as we've previously argued in \cite{Paulos:2019gtx}.

The master functionals that we have constructed in 1dare dependent on the value of $\Df$, the dimension of the external operator in the correlator, whereas this is not the case in the higher dimensional version. In fact the higher dimensional functionals take a much simpler form than the 1d ones. The price to pay is that the higher dimensional basis only makes manifest crossing symmetry in the $s,t$ channels. For instance, the Polyakov-Regge blocks of \cite{Mazac:2019shk} (see also \cite{Sleight:2019ive}) include odd-spins in their conformal block decomposition. Since in a CFT correlator of identical scalar fields odd-spin contributions drop out, this means that if we want to focus on such correlators there is a large ambiguity in the functional basis, since we are allowed to essentially add and subtract contributions from odd spin functionals. In other words, there are many possible choices of extremal functional basis which will behave in the desired way when acting on double trace states of {\em even} spin.
Another, related, set of ambiguities, follows from the fact there can be no functional basis which is fully dual to generalized free bosonic fields. This is because this solution will in general admit Regge bounded deformations (such as AdS contact terms) which must be bootstrappable by any hypothetical functional basis. 

Overall, this means that there is a lot of freedom in how to define Polyakov-Regge blocks and it is not clear which choice is best. These ambiguities have important consequences for a bootstrapper, since they can in principle completely modify the positivity properties of the Polyakov-Regge blocks and therefore of the extremal functional basis, positivity properties which are crucial for obtaining bounds.

\paragraph{Upper bound on CFT correlators?} 
Our final and most important point, which is related to the ones we have already made and which we will now examine in detail, is that the dispersion relations of \cite{Carmi:2019cub,Caron-Huot:2020adz} necessarily require subtractions. These subtractions were of course also necessary for the dispersion relations that we have discussed in this work. The major difference however is that in the higher dimensional case such subtractions are typically infinite in number. These subtractions do not have definite signs and this spoils many possible bounds which otherwise might be derived. For a crossing symmetric CFT correlator the higher dimensional dispersion relations can be written in the form
\bea
\overline{\mathcal G}(w,\bar w)=-\int_{-\infty}^0\ud z\int_{-\infty}^0 \ud \bar z\, K_{w,\bar w}(z,\bar z) \mbox{dDisc$_s$} \overline{\mathcal G}(z,\bar z)
\eea
We will not need the precise form of $K_{w,\bar w}$ here, it is sufficient to say that it is the sum of two pieces which restrict the domain of integration and which furthermore are {\em positive} in that domain. In fact there are several possible choices of $K_{w,\bar w}$, but here we have in mind the subtracted kernel proposed in section 4.3 of \cite{Caron-Huot:2020adz}. 

The above equation is valid only for a suitably subtracted correlator. This is because the region of integration probes the lightcone and double lightcone OPE of the correlator. For instance we have
\bea
K_{w,\bar w}(z,\bar z)\underset{z\to 0}{=} O(z^{-1})
\eea
In the same limit conformal blocks behave as $z^{\frac{\tau}2-\Df}$ and therefore it follows that we must subtract out at least all operators in $\mathcal G$ which have twist $\tau\leq 2\Df$. In practice we choose to subtract out crossing-symmetric (in $s$ and $t$ channel) Polyakov-Regge blocks
\bea
\overline{\mathcal G}(z,\bar z):= \mathcal G(z,\bar z)-1-\sum_{0\leq \tau\leq 2\Df} a_{\Delta,\ell} \mathcal P_{\Delta,\ell}(z,\bar z)\,.
\eea
The Polyakov blocks here are defined as the action of the dispersion relation on a single conformal block, suitably analytically continued to the region $\tau\leq 2\Df$. They are crossing symmetric with respect to the $s$ and $t$ channel OPEs, but not $u$ channel. Again we point out that it is not clear whether these subtractions are sufficient, since the integral in the dispersion relation also probes the double lightcone limit $z\to 0^-,\bar z \to -\infty$ where the singularity of the correlator may be stronger. For the sake of argument we will assume we are working in a class of correlators where such subtractions are not necessary.

The dispersion relation implies the bound
\bea
\overline{\mathcal G}(z,\bar z)\leq 0\,,\qquad \mbox{for}\quad u=z\bar z\geq 0\,, v=(1-z)(1-\bar z)\geq 0
\eea
thanks to the positivy of the kernel and that of conformal blocks in that domain. Expanding out this means
\bea
\mathcal G(z,\bar z)\leq \mathcal G_{\mbox{\tiny gff}}(z,\bar z)+\hat \sum_{\tau\leq 2\Df} a_{\Delta,\ell} \mathcal P_{\Delta,\ell}(z,\bar z)
\eea
where the hatted sum means we are not considering the contribution of the identity operator, having explicitly separated it out:
\bea
1+\mathcal P_{0,0}(z,\bar z)=1+\frac{1}{u^{\Df}}+\frac{1}{v^{\Df}}=\mathcal G_{\mbox{\tiny gff}}(z,\bar z)\,.
\eea
This is tantalisingly close to the upper bound derived in $d=1$ and given in equation \reef{eq:corrbounds}. However we see here that the subtractions are in the twist, not scaling dimension. In particular, setting a gap $\Delta_g=2\Df$ in the spin $\ell=0$ sector of the CFT still leaves out a possible infinite number of subtractions whose sign is not determined and must be investigated by hand. In particular if those subtractions would have a negative sign we would be done. This can be investigated using the dispersion relation. We set $w=\bar w$ and $w\in(0,1)$ in which case we have
\bea
\mathcal P_{\Delta,\ell}(w,w)=-2\sin^2\left[\frac{\pi}2(\tau-2\Df)\right]\,\int_0^1 \ud z \int_{z}^1 \ud \bar z\, K_{w,\bar w}\left(\mbox{$\frac{z}{z-1},\frac{\bar z}{\bar z-1}$}\right)\,\frac{G_{\Delta,\ell}(z,\bar z)}{\left(\frac{z}{1-z}\frac{\bar z}{1-\bar z}\right)^{\Df}}
\eea
In the $z\to 0$ limit the integral can develop a singularity for $\tau<2\Df$ due to power law behaviour of the integrand of the form $z^{\frac{\tau}2-\Df-1}$. It can be checked that the $\bar z$ integral does not change the nature of this singularity. This divergence changes the double zero of the Polyakov block at $\tau=2\Df$ into a simple zero and hence the sign of the block flips beyond this point. In general the Polyakov block can be computed for smaller values of twist by a subtraction trick. The upshot is that we are able to compute the Polyakov block for $\tau\leq 2 \Df$ and check that it can be positive. 

Hence as it stands unfortunately no bound on the correlator exists unless we impose a gap on the twist. Following our previous discussion, we believe that it should be possible to fix this by redefining the Polyakov blocks to improve their positivity properties. How this could be done remains to be understood.

\subsection{Summary and Outlook}

In this paper we have introduced a new way of thinking about the constraints of crossing symmetry on the line $z=\bar z$. We have seen that the 1d functional bases are neatly captured by master functionals, which imply validity of the Polyakov bootstrap as well as certain crossing-symmetric dispersion relations for CFT correlators. The dispersion relations fully capture the constraints of crossing symmetry, imply new universal bounds on the values of CFT correlators and allow us to efficiently bootstrap them in terms of their double discontinuity. They can also be used to probe and constrain the Regge limit of higher-$d$ CFT correlators.  

Our study of the Regge limit is closely related to the study of the late time limit of out-of-time-order correlators (OTOC) . From a $d=1$ perspective this limit is given by\footnote{This limit was studied from a different perspective in \cite{Ferrero:2019luz}.}
\bea
\mathcal G(z)|_{z=\frac 12+i t} \underset{t \to \infty}{\sim} \zeta\, t^{r_0-1}\,.
\eea
Having $r_0<1$ is then a potential signal that the underlying system is chaotic \cite{Roberts:2014ifa}.
Our analysis in section \ref{sec:regge} is useful also for determining the value of $r_0$. Concretely \reef{eq:largew} tells us that:
\bea
r_0=\zeta=1\qquad \mbox{unless}\qquad \tj=j_0=-i\mu=\nu=1
\eea
Thus we see that decaying correlators are highly constrained objects. More generally if $r_0<1$ then it can be determined by studying subleading terms in the expansion of the double discontinuity. Suppose for instance:
\bea
(1-z)^{2\Df}\mbox{dDisc}\, \mathcal G(z) \underset{z\to 1} \sim 1+ \nu_1 (1-z)^{1-\bar \jmath_1}+\ldots \label{eq:ddisc1}
\eea
In this case we would have $r_0=\mbox{max}\left\{1-\Delta_g,\bar \jmath_1\right\}$, unless there are further cancellations. It would be interesting to understand this in more detail. 

We've also argued the behaviour of $\tj$ is related to the properties of high dimension operators. Decay of the correlator thus requires that the average anomalous dimensions are order one. One extreme for getting this is a chaotic spectrum where at high energies we get essentially a continuum of states. Another, it seems, is highly organized spectra. This is for instance what happens for the 2d Ising correlator
\bea
\mathcal G_{\sigma \sigma \sigma \sigma}(z,z)=\frac{1}{z^{2\Delta_{\sigma}} (1-z)^{2\Delta_{\sigma}}} \underset{t\to \infty}{\sim} \frac{1}{t^{4\Delta_{\sigma}}}\,, \qquad \Delta_{\sigma}=1/8\,,
\eea
for which the double discontinuity takes the form \reef{eq:ddisc1} with $\bar \jmath_1=4\Delta_\sigma$. In this case there are two towers of states with dimensions $2+2n$ and $1+2n$ and $\overline{\gamma_n^2}\underset{n\to \infty}=O(1)$.

A possible outcome of the present work is a shift in our philosophy for obtaining constraints out of the crossing equation. There is an ongoing program of constructing bases of extremal functionals in various setups. But we have seen that the constraints which arise from these can be fully equivalent to certain positive dispersion relations. Importantly, these dispersion relations are in fact so constrained that they can be fixed uniquely: this was what we used in section \ref{sec:dispderivation} to rederive the master-functional kernels, without actually ever making reference to functionals at all! It seems to us that this kind of logic might be extremely useful for obtaining functional constraints on generic systems of CFT correlators. That is, one should start by {\em postulating} dispersion relations for a {\em matrix} of correlators of the schematic form
\bea
\mathcal G_{m}(w)=\int_0^1 \ud z g_{m}^{n}(w,z)\, \mbox{dDisc}\, \mathcal G_{n}(z)\,.
\eea
with $m,n$ multi-indices and impose
\bea
\mbox{dDisc}\, g_{m}^n(w,z)=\delta(w-z) \delta_m^n\,, \qquad g_{m}^n(1-w,z)=C_m^p(w) g_{p}^n(w,z)
\eea
with $C_m^p$ encoding crossing symmetry relations. These equations could then be used to completely fix $g_m^n$ under suitable boundary conditions. Positive semi-definitess of the matrix of CFT correlators should imply similar properties on the kernels, leading to functionals with good positivity properties. It would be extremely interesting to pursue this and to see if a similar strategy can be made to work in higher dimensions. 

The dispersion relations that we have found can be studied in the limit of large $\Df$\footnote{Work in progress with L. C\'ordova and Y. He.}. Such a limit is interesting as it important for understanding CFT correlators associated to massive QFTs in AdS$_2$ \cite{Paulos:2016fap}. In turn such correlators are closely related to S-matrices \cite{Dubovsky:2017cnj,Komatsu:2020sag}. The dispersion relation can then be used to study analyticity properties of 2d S-matrices. An analysis in this limit shows that there is a one-to-one mapping between extremal solutions to crossing/S-matrices and dispersion relations with nice positivity roperties. Perhaps this relation can be used to improve our understanding of non-trivial interacting solutions to crossing at finite $\Df$.

Finally, the ideas introduced in this work have interesting applications for the numerical bootstrap. Minimizing or maximizing CFT correlator values provide new ways to probe the boundaries of the CFT landscape. Perhaps interesting CFTs which have up to now escaped the probings of bootstrappers will be found this way. In the context of this work it was crucial that at least sometimes these optimization problems reduce to rather well known ones: gap maximization and OPE maximization. This is not always the case in more general settings, but even when it is it can be useful: minimizing a CFT correlator value can be done in a single optimization step, while the gap maximization problem requires a computationally costly binary search. We hope to report on explorations of these topics soon \cite{toappearZechuan}.

\section*{Acknowledgments}
This work has benefitted from discussions with L. C\'ordova, Y. He, D. Maz\'a\v{c}, J. Penedones, S. Caron-Huot and especially Z. Zheng.
We would like to thank the Simons Collaboration on the Nonperturbative Bootstrap for leading to opportunities for discussion and collaboration.
\pagebreak

\appendix

\section{Computing functional kernels} 
\label{app:funcsol}
In this appendix we will discuss various methods for computing the master functional kernels.

\subsection{Kernels for particular cases}

One way to find the functional kernels is to directly solve 
equations \reef{eq:fundfreeeq} subject to the boundary conditions \reef{eq:boundcond}. In general this is difficult, but simplifications occur for special values of $\Df$. For the $\OL_w$ functional and $\Df$ half-integer, or integer $\Df$ and $\OU_w$, we can find solutions by means of an ansatz. We write
\bea
\ful_w(z)=\frac{p_w(z) \log\left(\frac{z-1}z\right)+q_w(z)}{z(z-1)\prod_{w^T\in \mathcal T}(w^T-z)}\,,\qquad \mathcal T=\left\{w,1-w,\frac{1}{1-w},\frac{1}w,\frac{w}{w-1},\frac{w-1}w\right\}
\eea
with $p_w(z)=-p_w(1-z)$ and $q_w(z)=q_w(1-z)$ polynomials in $z$ of some fixed degree.  Plugging in such an ansatz into equation \reef{eq:fundfreeeq} we can find solutions for the special cases mentioned above. The coefficients in such polynomials become functions of $w$ fixed by equation 
\reef{eq:fundfreeeq} as well as by the absence of singularities in $z$ when $|z|>1$.
As an example we find:
\bea
\Df=\frac 12:\quad \fl_w(z)=-\frac{1}{\pi^2}\left(\frac{1}{z-\frac{1}{w}}+\frac{1}{-\frac{1}{w}-z+1}\right) \left(\frac{(2 z-1) \log \left(\frac{z-1}{z}\right)}{2-w}
-\frac{\log
   (1-w)}{w}\right)\\-\frac{1}{\pi^2}\left(\frac{1}{z-\frac{1}{1-w}}+\frac{1}{-\frac{1}{1-w}-z+1}\right) \left(\frac{(2 z-1) \log
   \left(\frac{z-1}{z}\right)}{w+1}-\frac{\log (w)}{1-w}\right)+\\
	-\frac{1}{\pi^2}\left(\frac{1}{w+z-1}+\frac{1}{w-z}\right) \left(\frac{(2 z-1) \log
   \left(\frac{z-1}{z}\right)}{2 w-1}+\log \left(\frac{w}{1-w}\right)\right) \label{eq:examplef}
\eea
It is straightforward to find other solutions for increasing $\Df$ although the corresponding expressions become increasingly absurd. In all cases that we have been able to construct the kernels, we always find that the $\hat g$ kernels are positive for $z\in(0,1)$.

Another case where we are able to explicitly find the kernels is in the limit of large $\Df$. Assuming a finite limit for $f_w(z)$ the fundamental equation \reef{eq:fundfreeeq} simplifies to
\bea
\mathcal R_z \fl_w(z)&\underset{\Df\to \infty}\sim\delta(z-w)+\delta(1-z-w)\\
\mathcal R_z \fu_w(z)&\underset{\Df\to \infty}\sim -\delta(z-w)-\delta(1-z-w)
\eea
The correct solutions satisfying boundary conditions \reef{eq:boundcond} are given by
\bea
\fl_w(z)=-\fu_w(z)=\frac{2}{\pi}\sqrt{\frac{w(1-w)}{z(z-1)}} \frac{\frac 12-z}{(w-z)(1-w-z)}
\eea
which also lead to positive $\ghul_w(z)$ (cf. footnote \ref{fn:zz}).
Actually, the boundary condition near $z=0$ cannot be satisfied exactly, since the small $z$ limit does not commute with that of large $\Df$. Nevertheless the small $z$ behaviour of the solution above is softer than that of any homogeneous solution and this requirement still fixes it uniquely.

\subsection{Computing from basis expansion}

We will now show how to compute functional kernels by starting directly from the basis expansions \reef{eq:sumexprl} and \reef{eq:sumexpru}. For the purposes of this subsection, we will work with the functional definition given in \reef{eq:funch}:
\bea
\omega[\mathcal F]=\int_1^{\infty} \frac{\ud z}{\pi} h_\omega(z) \mathcal I_z \mathcal F(z).
\eea
The kernels associated to the master functionals $
OUL_w$ given in \reef{eq:sumexprl}, \reef{eq:sumexpru}are then
\bea
h_w^F(z)&=-\sum_{n=0}^{\infty}\left[ G_{\Delta_n^F}(w|\Df) h_{\alpha_n^F}(z)+\partial_{\Delta}G_{\Delta_n^F}(w|\Df) h_{\beta_n^F}(z)\right]\,,\\
h_w^B(z)&=+\sum_{n=0}^{\infty}\left[ G_{\Delta_n^B}(w|\Df) h_{\alpha_n^B}(z)+\partial_{\Delta}G_{\Delta_n^B}(w|\Df) h_{\beta_n^B}(z)\right] \label{eq:kernelseries}
\eea
To determine the lefthand side we need to know the kernels of the bosonic and fermionic basis. These are in principle known for the special cases where $\Df$ is an integer/half-integer respectively but have not yet been worked out in detail in the literature. Here we will only present the special case corresponding to the kernel $h_w^F$ and $\Df=1/2$. Our starting point are the kernels for the fermionic basis\footnote{Some of these have appeared in \cite{Mazac:2016qev}.}:
\bea
h_{\beta_n^F}(z)&=\frac{2}{\pi^2} \frac{\Gamma(\Delta_n^F)^4}{\Gamma(2\Delta_n^F)\Gamma(2\Delta_n^F-1)}\,\left(\frac{G_{\Delta_n^F}(\mbox{$\frac{z-1}z$})}{z-1}+G_{\Delta_n^F}(\mbox{$\frac{1}z$})-\frac{Q_{\Delta_n^F-1}\left(\mbox{$\frac{z-2}z$}\right)}{z}\right)\\
h_{\alpha_n^F}(z)&=\frac{1}{2}\partial_n h_{\beta_n^F}(z)
\eea
with $\Delta_n^F=1+2\Df+2n=2+2n$ and
\bea
Q_{\Delta-1}\left(\mbox{$\frac{z-2}z$}\right):=\mathcal R_z G_{\Delta}\left(\mbox{$\frac{z}{z-1}$}\right)\,, \qquad z>1.
\eea
From the expressions it is clear that the desired result will be obtained if we can first compute
\bea
p_{w}^F(z)=\frac{1}{\pi^2} \sum_{n=0}^\infty\partial_n \left[\frac{\Gamma(\Delta_n^F)^4}{\Gamma(2\Delta_n^F)\Gamma(2\Delta_n^F-1)}\,\frac{G_{\Delta_n^F}(w) G_{\Delta_n^F}(z)}{w z}\right]
\eea
Let us ignore the derivative. We can write
\bea
\frac{\Gamma(\Delta_n^F)^2}{\Gamma(2\Delta_n^F)}\frac{G_{\Delta_n^F}(w)}w=\int_1^{\infty}\frac{\ud x}{\pi} \frac{P_{\Delta_n^F-1}(\frac{2-x}x)}{x(x-w)}
\eea
where $P_n(z)$ is the $n$-th Legendre polynomial. We can now use the completeness relation of those polynomials,
\bea
\sum_{n=0}^\infty (2n+1) P_{n}(z)P_n(w)=\delta(z-w)
\eea
integrating twice and antisymmetrizing to find
\begin{multline}
\frac{2}{\pi^2} \sum_{n=0}^\infty \left[\frac{\Gamma(\Delta_n^F)^4}{\Gamma(2\Delta_n^F)\Gamma(2\Delta_n^F-1)}\,\frac{G_{\Delta_n^F}(w) G_{\Delta_n^F}(z)}{w z}\right]\\=\frac{(2-w) z \log(1-w)- (2-z) w \log(1-z)}{\pi^2 (w-z)[z(w-1)-w]}
\end{multline}
This is all well and good but still not the result we want, since we must account for the derivative. When it acts on the conformal blocks it leads to terms that behave like $\log(w)$ and $\log(z)$ which will multiply the result above. So we need to find some function whose discontinuity for negative $z$ and $w$ matches the above. The crucial (if fortuitous) observation is that the above can be written as
\bea
\frac{(2-w) z \log(1-w)- (2-z) w \log(1-z)}{\pi^2 (w-z)[z(w-1)-w]}=-\frac{G^{4d}_{\Delta=3,\ell=1}(z,w)}{3\pi^2[z(w-1)-w]}\label{eq:pp}
\eea
where $G_{\Delta,\ell}$ stands for the 4d conformal block! Hence we guess, correctly as it turns out:
\bea
p^F_w(z)=\frac{2}{\pi^2}\frac{\partial_{\Delta} G^{4d}_{\Delta=3,\ell=1}(z,w)-\frac 43\, G^{4d}_{\Delta=3,\ell=1}(z,w)}{3[w-z(w-1)]}\,.
\eea
This may be computed explicitly using:\footnote{The 4d conformal blocks have known expressions in terms of products of 1d conformal blocks \cite{Dolan:2000ut}. The latter and their derivatives with respect to $\Delta$ satisfy recursion relations for integer arguments (which follow from those of Legendre polynomials) which can be used to determine them.}
\begin{multline}
\frac 23 (w-z)\left[\partial_{\Delta} G^{4d}_{\Delta=3,\ell=1}(z,w)-\frac 43\, G^{4d}_{\Delta=3,\ell=1}(z,w)\right]
=2 (w-2) z \text{Li}_2(w)-2 w (z-2) \text{Li}_2(z)\\-w (z-2) \log (1-z) (\log (w)+\log (z))\\+\log (1-w) ((w-2) z \log (w)+(z-w) \log
   (1-z)+(w-2) z \log (z))\,.
	\end{multline}
We can now get
\bea
h^F_w(z)=\frac{1}z\,\left[p^F_w(\mbox{$\frac{z-1}z$})+p^F_w(\mbox{$\frac 1z$})\right]-\mathcal R_z p^F_w(z)\,, \qquad w\in (0,1)\,,\quad z>1
\eea
The complete expression is relatively simple but longish, so we will not write it out in full here. However, it is easy to check using \reef{eq:fgfromh} that it leads to the same kernel $\fl_w$ as in \reef{eq:examplef}. Thus we have shown that the two definitions of the functional lead to the same result, at least for this particular case.

What about $\gl_w$? Recall that this is given as a sum of a delta function plus an analytic piece in $z\in (0,1)$. Does the above kernel correctly recover the $\delta$ function piece? Indeed it does as it can be checked explicitly. For instance, quite non-trivially one finds that the above kernel satisfies the identity:
\bea
h^F_w(z)-h^F_{1-w}(z)=-\frac{1-2w}{(z-w)(1-z-w)}\,.
\eea
This is nothing but the statement at the level of the $h$ kernel of the property
\bea
\OL_w-\OL_{1-w}=-\mathcal E_w
\eea

\subsection{General $\Df$ and contact term expansion}
\label{app:contacts}
An interesting way to think about the kernels $\ful_w(z)$ are as generating functions for contact interactions in AdS. For definiteness let us focus on the bosonic master functional. For our purposes AdS contact interactions amount to certain solutions of the crossing equations of the form
\bea
\mathcal C^B_k(w)=\mathcal C^B_k(1-w) \Leftrightarrow \sum_{n=0}^\infty \left[c_n^{(k)} F_{\Delta_n^B}(z)+d_n^{(k)} \partial_{\Delta} F_{\Delta_n^B}(z)\right]=0\,.\label{eq:crosscontact}
\eea
There are an infinite set of solutions of this form  parameterized by the Regge or large-$w$ behaviour of $\mathcal G^B_k$ which is of the form $w^{2k-1}$ with $k\geq 0$ integer (or equivalently by the growth in $n$ of the coefficients $c_n^{(k)}, d_n^{(k)}$). There is clearly an ambiguity in the choice of the basis of contact terms which we will fix by the requirement
\bea
d_n^{(k)}=\delta_{n,k} \qquad \mbox{for}\quad n\leq k\,.
\eea 
Note that applying the bosonic basis functionals to the crossing equations above we identify $c_n$ and $d_n$ in \reef{eq:dualityboson} with $c_n^{(0)}$ and $d_n^{(0)}$ respectively. Why? Well here $k=0$ necessarily since such functionals can only be applied to correlators which behave as $z^{1-\epsilon}$ for large $z$. But suppose we wanted to have functionals which could bootstrap solutions with higher $k$. To do that we need to construct functionals whose $f(z)$ kernels have softer behaviour at large $z$. Set $\alpha_n^{B,(0)}=\alpha_n^B, \beta_n^{B,(0)}=\beta_n^B$ and define recursively:
\bea
\alpha_n^{B,(k+1)}:=\alpha_n^{B,(k)} - c_n^{(k+1)} \beta_{k+1}^{B,(k)}\,, \qquad \beta_n^{B,(k+1)}:=\beta_n^{B,(k)} - d_n^{(k+1)} \beta_{k+1}^{B,(k)}\,, \qquad k\geq 0
\eea
Note in particular that $\beta_n^{B,(k)}=0$ for $0\leq n \leq k$. The coefficients $c_j^{(k)}$ and $d_j^{(k)}$ are not only those appearing in the expressions for the contact terms, they are also precisely those for which the functional kernels $f_{\omega^{B,(k)}}(z)$ decay as $1/z^{2+2k}$. Indeed we can apply such functionals to bootstrap contact terms $\mathcal C_{k'}^B$ for $k'=0,1,\ldots, k$, and with our definition of these it is easy to check this matching must be true.

Let us go back to the bosonic master functional and consider
\bea
\OU_w-\sum_{j=0}^{k-1} \mathcal C^B_{j+1}(w) \beta_{j+1}^{B,(j)}=\sum_{n=0}^\infty\left[ G_{\Delta_n^B}(w|\Df) \alpha_n^{B,(k)}+\partial_\Delta G_{\Delta_n^B}(w|\Df)\beta_n^{B,(k)}\right]
\eea
At the level of the functional kernels, every one appearing on the righthand side decays at large $z$ as $z^{-2-2k}$. 
It follows then that 
\bea
\fu_w(z)\underset{z\to \infty}{\sim}\sum_{j=0}^{k} \mathcal C^B_{j+1}(w) f_{\beta_{j+1}^{B,(j)}}(z) +O(z^{-4-2k})\,.
\eea
We thus see that the large $z$ expansion of $\fu_w(z)$ nicely encodes almost the entire set of contact interactions. A similar result holds for the $\fl_w$ kernel. We can use this result both ways. If we know $\fu_w(z)$ we can expand it at large $z$ to obtain contact terms. Conversely, if we know how to compute contact terms we can combine this with our knowledge of the basis functional kernels to determine $\fu_w(z)$ in a large-$z$ expansion.

\section{Asymptotics of Polyakov blocks}
\label{app:asymp}
We are interested in understanding the large $w$ limit of the Polyakov blocks $\mathcal P^{B,F}(w)$. We will focus on the bosonic block the discussion being analogous for the other case. Let us write
\bea
\mathcal P_{\Delta}^B(w|\Df)=G_{\Delta}(w|\Df)-\OU_w(\Delta)
\eea
At large $w$ the second term dominates. According to \reef{eq:funcactionfg} the functional action contains two pieces. We cannot expand the kernels in powers of $w$ directly since the integral might then diverge.
In particular from the $g$ piece we have
\bea
\hat g_w^B(z) F_{\Delta}(z|\Df) \underset{z\to 1, w\to \infty}{\sim}\frac{2}\pi \frac{i \,w}{(w-\frac 1{1-z})(w+\frac{1}{1-z})}\, (1-z)^{\Delta-2}\,, \qquad  \quad \mbox{for fixed}\quad (1-z)w\,.
\eea
We cannot expand this term inside the integral sign unless $\Delta>1$. In any case this gives a contribution $\kappa(\Delta) (1-w)^{-\Delta}$ in the large $w$ limit. If $\Delta>1$ we can expand the $\fu_w, \gu_w$ kernels to leading order in $w$\,,
\bea
\fu_w(z) \underset{w\to \infty}{\sim} \frac{i \pi}{w} f_{\tilde \beta_0^B}(z)
\eea
Here $f_{\tilde \beta_0^B}$ is the functional kernel associated to the functional $\tilde \beta_0^B$. We have checked this in all cases where we have analytic expressions for $\fu_w(z)$. More generally, this can be shown from the following argument. 

The $\tilde \beta_0$ functional is not crossing compatible (it is $O(1)$ as $z\to \infty)$ and satisfies the duality conditions
\bea
\tilde \beta_0^B(\Delta_n^B)=0\,,\qquad \tilde \partial_\Delta\beta_0^B(\Delta_n^B)=\delta_{n,0}\,.
\eea
In fact this is but one member of a family of functionals with the same kind of fall off,
\bea
\tilde \alpha_n^B(\Delta_m^B)&=\delta_{n,m},&\qquad \partial_{\Delta} \tilde \alpha_n^B(\Delta_m^F)&=0,\\
\tilde \beta_n^F(\Delta_m^B)&=0,&\qquad  \partial_{\Delta}\tilde \beta_n^F(\Delta_m^B)&=\delta_{n,m}\,,\label{eq:dualityboson2}
\eea
which should be contrasted with \reef{eq:dualityboson} for ordinary functionals. These functionals bootstrap a different set of Polyakov blocks $\widetilde{\mathcal P}^B_{\Delta}$ which decay at large $w$ as $w^{-3}$ \cite{Mazac2019a}. We have
\bea
\mathcal P^B_{\Delta}(w)=\widetilde{\mathcal P}_{\Delta}^B-\tilde \beta_0(\Delta)\, \mathcal C_0(w) \label{eq:polydiff}
\eea
where $\mathcal C_0(w)$ is the unique AdS$_2$ contact term decaying as $1/w$ for large $w$, see section \ref{app:contacts}\,. This follows from
\bea
\alpha_n^B=\tilde \alpha_n^B-c_n \tilde \beta_0^B\,,\qquad \beta_n^B=\tilde \beta_n^B-d_n \tilde \beta_0^B\,.
\eea
It follows from \reef{eq:polydiff} that 
\bea
\mathcal P^B_{\Delta}(w) \underset{w\to \infty}\supset\frac{ -i \pi}{w} \tilde \beta_0(\Delta)
\eea
Overall we conclude:
\bea
\mathcal P^B_{\Delta}(w) \underset{w\to \infty}\sim \frac{\kappa(\Delta)}{(-w)^{\Delta}}-i \pi \frac{\tilde \beta_0(\Delta)}{w}
\eea
which is the desired result.

\small
\parskip=-10pt
\newpage
\bibliography{biblio}
\bibliographystyle{jhep}

\end{document}